\documentclass[twocolumn,aps,prb,nofootinbib]{revtex4-2}
\usepackage{graphicx}
\usepackage{times}
\usepackage{bm}
\usepackage[linktocpage=true,colorlinks=true,urlcolor=blue,linkcolor=red,citecolor = blue]{hyperref}
\usepackage{pgfplots}
\usepackage{amsfonts}
\usepackage{amsmath}
\usepackage{amssymb}
\usepackage{xcolor}
\usepackage{braket}
\usepackage{relsize}
\usetikzlibrary{arrows}
\usepackage{enumerate}
\usetikzlibrary[patterns] 
\usetikzlibrary[snakes] 
\usetikzlibrary[shapes] 
\usepackage{pgfplots}
\usepackage{newfloat}
\usepackage{siunitx}
\usepackage{physics}
\usepackage{mathtools}

\newcommand{\printfnsymbol}[1]{%
  \textsuperscript{\@fnsymbol{#1}}%
}

\DeclareFloatingEnvironment[name={Supplementary Figure}]{suppfigure}

\graphicspath{ {Figures/} }

\begin{document}

\title{Enhanced topological superconductivity in spatially modulated planar Josephson junctions}
\author{Purna P. Paudel}
\author{Trey Cole}
\author{Benjamin D. Woods}
\email{bdw0010@mix.wvu.edu}
\author{Tudor D. Stanescu}
\affiliation{Department of Physics and Astronomy, West Virginia University, Morgantown, WV 26506, USA}

\begin{abstract}
We propose a semiconductor-superconductor hybrid device for realizing topological superconductivity and Majorana zero modes  consisting of a planar Josephson junction structure with periodically modulated junction width. By performing a numerical analysis of the effective model describing the low-energy physics of the hybrid structure, we demonstrate that the modulation of the junction width results in a substantial enhancement of the topological gap and, consequently, of the robustness of the topological superconducting phase and associated Majorana zero modes. This enhancement is due to the formation of minibands with strongly renormalized effective parameters, including stronger spin-orbit coupling, generated by the effective periodic potential induced by the modulated structure. In addition to a larger topological gap, the proposed device supports a topological superconducting phase that covers a significant fraction of the parameter space, including the low Zeeman field regime, in the absence of a superconducting phase difference across the junction. Furthermore, the optimal regime for operating the device can be conveniently accessed by tuning the potential in the junction region using, for example, a top gate. 
\end{abstract}

\maketitle

\section{Introduction}

Majorana zero modes (MZMs) are particle-hole symmetric zero-energy excitations emerging  in topological superconductors, either at the edges of a one-dimensional (1D) system, or inside the vortex cores of a two-dimensional (2D) system \cite{Bravyi2002}. Their non-Abelian exchange statistics makes them attractive for quantum computation, as possible building blocks for topological qubits \cite{Nayak2008, Alicea2011}. Such qubits are, in principle, immune to decoherence from local perturbations due to the non-local encoding of quantum information using pairs of spatially-separated MZMs \cite{Kitaev2003}. A major breakthrough was generated by the proposal to engineer topological superconductors capable of supporting MZMs using semiconductor-superconductor (SM-SC) hybrid nanostructures \cite{Lutchyn2010,Oreg2010,Sau2010,Sau2010a}. This  has sparked a significant experimental effort over the last decade, which lead to tremendous progress in the area of materials growth and device engineering and generated a number of observations consistent with the presence of MZMs \cite{Mourik2012,Deng2012,Das2012,Churchill2013,Finck2013,Albrecht2016,Chen2017,Gul2017,Deng2018,Vaitiekenas2018,Chen2019,Grivnin2019,Vaitiekenaseaav2020,Yu2021,Zhang2021}. However, there is an ongoing debate  within the community regarding the nature of the low-energy modes responsible for the experimental observations, as trivial  low-energy Andreev bound states emerging in the presence of disorder or system inhomogeneity are capable to mimick the (local) Majorana phenomenology \cite{Kells2012,Bagrets2012,Liu2012,Prada2012,Roy2013,Adagideli2014,Cayao2015,Liu2017a,Reeg2018a,Moore2018,Moore2018a,Woods2019b,Pan2020,Prada2020,Woods2021,Pan2021a,Sarma2021}. This rather uncertain situation has motivated the exploration of alternative paths for realizing MZMs, including new designs of SM-SC heterostructures. 

One interesting alternative to the ``standard'' Majorana nanowire platform consists of planar SM-SC structures, which have attracted both experimental \cite{Shabani2016,Kjaergaard2016,Nichele2017,Suominen2017,OFarrell2018,Fornieri2019,Lee2019a,Ren2019,Pankratova2020,Mayer2020,Dartiailh2021} and theoretical  \cite{Hell2017a,Pientka2017,Hell2017,Reeg2018,Setiawan2019a,Setiawan2019b,Scharf2019,Melo2019,Woods2020b,Laeven2020,Zhang2020,Zhou2020,Pakizer2021,Lesser2021,Alidoust2021} attention in recent years. 
This type of  system involves a two-dimensional electron gas (2DEG) with strong spin-orbit coupling and large g-factor, which is hosted by a semiconductor hetersrostructure, in proximity to a superconducting film (often Al) and in the presence of an in-plane magnetic field. The SC is patterned either as  a narrow quasi-1D strip (nanowire), with the 2DEG outside the strip being depleted using a top gate  \cite{Nichele2017,Suominen2017,Woods2020b}, or as a Josephson junction consisting of two (large) superconducting regions separated by a narrow (unproximitized) channel \cite{Fornieri2019,Ren2019,Dartiailh2021,Hell2017,Pientka2017}. The Josephson junction design has an additional experimentally- tunable parameter: the superconducting phase difference $\phi$. As $\phi$ is tuned from $\phi = 0$ to $\phi = \pi$, the necessary Zeeman splitting $E_Z$ needed to drive the system into the topological phase can approach zero \cite{Hell2017a,Pientka2017},  which is a very attractive feature. Moreover, for $\phi = \pi$  accessing the topological phase does not require  fine-tuning the chemical potential $\mu$, in stark contrast to the nanowire design, in which $\mu$ must be tuned near the bottom of a (confinement-induced) subband. Finally,  the planar designs are easily scalable to large networks of Majorana qubits, which will be necessary for quantum computation \cite{Divincenzo2000}. Majorana nanowires, on the other hand,  require the realization of exotic  networks \cite{Krizek2018} to become scalable. 

Despite these attractive features, planar SM-SC structures have several potentially serious  issues. For the SC strip design,  a dramatic suppression of the effective g-factor and spin-orbit coupling may occur if the coupling between the semiconductor and superconductor is strong \cite{Stanescu2017a,Reeg2018a}. In the strong coupling regime, the Zeeman energy necessary to reach the topological phase is no longer $E_{Z,crit} = \sqrt{\mu^2 + \Delta_{ind}^2}$, as found in the minimal 1D Majorana model \cite{Lutchyn2010,Oreg2010,Sau2010,Sau2010a}, but rather $E_{Z,crit} = \sqrt{\mu^2 + \widetilde{\gamma}^2}$ \cite{Stanescu2017a}, where $\Delta_{ind}$ is the induced superconducting gap, $\mu$ is the chemical potential, and $\widetilde{\gamma}$ is the effective SM-SC coupling. Note that $\widetilde{\gamma}$ can be several times larger than $\Delta_{ind}$, which makes it difficult to reach the topological phase before superconductivity in the parent superconductor is destroyed by the applied magnetic field. Many widely used hybrid structures, which include both nanowires and planar strip-type systems, particularly those characterized by a hard (induced) SC gap, are likely  to be in a strong coupling regime since the induced gap is comparable to the parent gap, $\Delta_{ind} \sim \Delta_o$. 

The Josephson junction design overcomes the issues associated with strong SM-SC coupling, basically because the  ``bare'' semiconductor parameters are not renormalized by the superconductor within the junction region. Also, this design enables an additional, potentially useful tuning parameter: the superconducting phase difference.
However, the Josephson junction system has a major issue that could make these attractive features practically irrelevant: the optimal topological gap is rather small, which makes the topological superconducting phase (and the corresponding MZMs) very fragile. In addition, realizing the conditions consistent with the optimal topological gap is  nontrivial, as it involves a highly uniform effective potential throughout the system. More specifically, for realistic device parameters characterizing InAs/Al systems, the topological gap $\Delta_{\text{top}}$ is restricted to rather small fractions of the parent gap $\Delta_o$. Combined with the already small parent SC gap of Al, $\Delta_o \approx 0.2-0.3~\text{meV}$, this leads to a very fragile  topological phase that can be easily destroyed by  a small amount of disorder \cite{Pan2020,Woods2021,Pan2021a}. 
Moreover, using superconductors with a larger gap does not automatically fix the issue, since it involves an increase of the Zeeman energy necessary to achieve the optimal topological gap \cite{Pientka2017}. Finally, the uniformity requirement to be satisfied by the effective potential represents a nontrivial task in the presence of both proximitized and non-proximitized regions \cite{Stenger2018} due to the strong band bending  at the SM-SC interface \cite{Woods2018,Winkler2019}, as well as the proximity induced shift of the potential in the presence of finite SM-SC coupling \cite{Stanescu2011,Reeg2018a}. 

In this work we propose a planar JJ device  with periodically modulated junction  width (see Fig. \ref{FIG1}) as a possible solution to the challenges facing the ``standard'' JJ design. In essence, the modulation of the junction width generates a strong periodic potential, which, in turn, produces minibands with strongly renormalized effective parameters. On the one hand, in the presence of these minibands, the  topological phase supported by the hybrid structure with no superconducting phase difference expands significantly within the low-field regime.  Most importantly, the topological gap characterizing the low-field topological phase is substantially enhanced as compared to gap associated with the the uniform, non-modulated structure (with or without a phase difference). In turn, this enhances the robustness of the topological phase and the associated MZMs against disorder. This enhancement of the topological gap is, in essence, a direct result of the larger effective spin-orbit coupling that characterizes the minibands induced by the periodic (effective) potential. Note that this type of optimization of the effective parameters by controlling the geometric properties of a nanodevice opens a new and potentially fruitful route to creating materials by design. Finally, we note that the optimal regime for operating the proposed modulated structure involves the presence of an attractive potential in the junction region, which can be easily generated using a top gate. This bypasses any potential uniformity requirement that may apply to the non-modulated structure. In addition, the low-field topological phase can be accessed by tuning the top gate potential without the need to apply a nonzero superconducting phase difference. This could significantly simplify the scaling of these devices and the design of multiqubit architectures. Of course, the proposed structure inherits the advantages of the JJ design with respect to the issues  arising from strong SM-SC coupling, which can affect nanowire-type devices.

The remainder  of this work is organized as follows. In Sec. \ref{Model} we introduce the theoretical model used to investigate the properties of  the modulated JJ device and present the numerical techniques developed to efficiently solve the corresponding quantum mechanics problem.
In Sec. \ref{Results} we first summarize the key properties of the uniform JJ structure, emphasizing the key issue of the relatively small topological gap, then we calculate the corresponding properties of the proposed modulated structure and demonstrate that the periodic modulation of the junction width leads to a significant enhancement of the topological gap and, implicitly, to a a more robust topological phase.  Our concluding remarks are presented in Sec. \ref{Conclusion}.

\section{Modeling} \label{Model}

\begin{figure}[t]
\begin{center}
\includegraphics[width=0.48\textwidth]{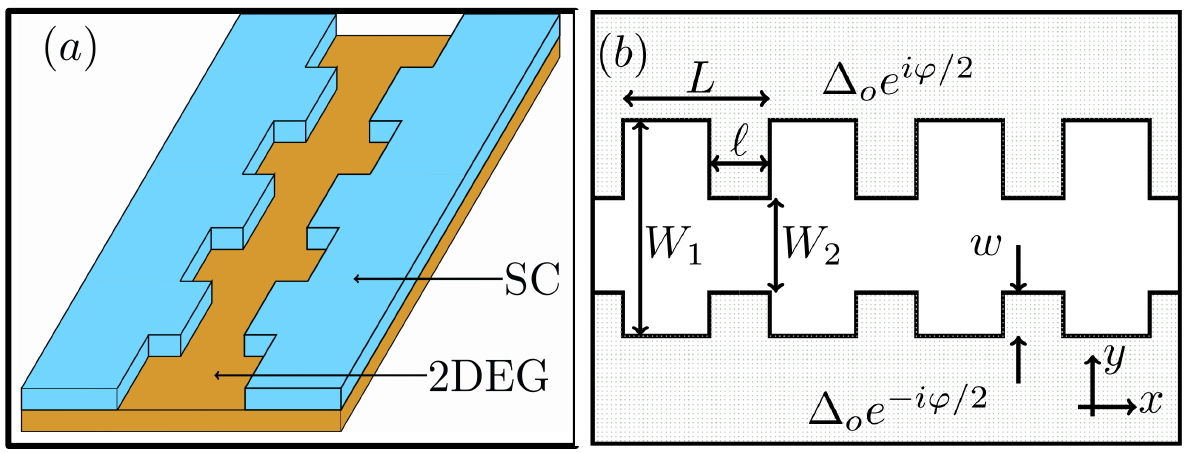}
\end{center}
\vspace{-2mm}
\caption{(a) Schematic representation of the proposed \textit{modulated} Josephson junction device. A 2DEG hosted by a semiconductor quantum well (orange) is proximity-coupled to a pair of s-wave superconductors (light blue) separated by a quasi-1D, spatially-modulated junction region. (b) Top view of four unit cells of the periodic structure. There is a phase difference $\phi$ between the two superconductors. The proximitized regions are considered to be semi-infinite in the $y$-direction.}
\label{FIG1}
\vspace{-1mm}
\end{figure}

We consider the hybrid system shown schematically in Fig. \ref{FIG1}, which consists of a 2DEG hosted by a semiconductor quantum well (InAs) that is proximity coupled to a pair of s-wave superconductors (Al) separated by a narrow quasi-1D region, forming a Josephson junction (JJ). The width of the JJ is periodically modulated, which represents the key new ingredient of our proposed design. Specifically, the width of the junction region varies between two values $W_1 \geq W_2$, where the constricted regions ($W_2$) have length $\ell$ and the overall length of the unit cell is $L$. In this section we discuss the theoretical model used to describe the system and the method developed to efficiently solve the corresponding quantum mechanics problem. We also provide a qualitative characterization of the effects generated by the periodic modulation of the junction width, which represent key ingredients affecting the low-energy physics of the proposed device.

\subsection{Effective low-energy Hamiltonian}

We assume that the quantum well hosting the 2DEG is narrow, so that its low-energy physics is accurately captured by a single quantized mode along the $z$-direction (i.e., the direction perpendicular to the 2DEG plane). We model the effectively two-dimensional semiconductor-superconductor (SM-SC) heterostructure at the mean-field level using a Bogliubov-de Gennes (BdG) Hamiltonian that, in the Nambu basis $(\psi_\uparrow, \psi_\downarrow, \psi_\uparrow^\dagger, \psi_\downarrow^\dagger)$, has the form
\begin{equation}
\begin{split}
    H =& 
    \left[
    -\frac{h^2}{2 m^*}\left(\partial_x^2 + \partial_y^2\right)
    - \mu + V(x,y)
    \right]\sigma_o\tau_z  \\
    &-i \alpha_R \Big(\partial_x\sigma_y \tau_z
    - \partial_y\sigma_x\tau_o\Big)
    + \widetilde{E}_Z(x,y) \sigma_x \tau_z \\
    &+ \Re\Big[ \Delta(x,y)\Big] \sigma_y \tau_y + \Im\Big[\Delta(x,y)\Big] \sigma_y \tau_x,
\end{split} \label{HBdG}
\end{equation}
where $m^*$ is the SM effective mass, $\mu$ is the chemical potential, $V$ is the electrostatic potential, $\alpha_R$ is the Rashba spin-orbit coefficient, $\widetilde{E}_Z$ is the Zeeman energy, $\Delta$ is the induced SC pairing, and $\sigma_j$ and $\tau_j$, with $j = o, x, y, z$, are Pauli matrices acting on the spin and particle-hole spaces, respectively. Note that $V$, $\widetilde{E}_Z$, and $\Delta$ are position dependent quantities having a periodic dependence on $x$ (i.e., on the position along the junction) with period $L$, $V(x + L,y) = V(x,y)$ and similarly for $\widetilde{E}_Z$ and $\Delta$. Explicitly, the electrostatic potential is assumed to have the form
\begin{equation}
    V(x,y) = 
    \begin{cases}
    V_J,\quad&(x,y)\in\text{junction region}, \\
    0,\quad&(x,y)\in\text{proximitized regions},
    \end{cases} \label{Pot}
\end{equation}
where $V_J$ is the junction potential. We note that the potential $V_J$ in the junction region can be controlled using a top gate, which will not affect the electrostatic potential of the proximitized regions due to screening by the SCs. 
The Zeeman energy $\widetilde{E}_Z$, which is generated by applying a magnetic field along the $x$-direction, has a spatial dependence given by
\begin{equation}
    \widetilde{E}_Z(x,y) = 
    \begin{cases}
    E_Z,\quad&(x,y)\in\text{junction region}, \\
    0,\quad&(x,y)\in\text{proximitized regions}.
    \end{cases} \label{Zeeman}
\end{equation}
Here, we assume that the junction region is characterized by a relatively large $g$-factor, while the effective $g$-factor within the proximitized regions is strongly suppressed as a result of the renormalization generated by a strong SM-SC coupling \cite{Stanescu2017a}. For simplicity, we neglect the small Zeeman splitting within the proximitized regions.
Finally, the induced SC pairing is assumed to have the form
\begin{equation}
    \Delta(x,y) = 
    \begin{cases}
    \Delta_o e^{i\phi/2},\quad&(x,y)\in\text{first SC region}, \\
    0,\quad&(x,y)\in\text{junction region}, \\
    \Delta_o e^{-i\phi/2},\quad&(x,y)\in\text{second SC region},
    \end{cases} \label{Delta}
\end{equation}
where $\Delta_o \in \mathbb{R}^+_0$ is the magnitude of the induced SC gap in the proximitized regions and $\phi$ is the phase difference between the two semi-infinite superconductors, which extend along the y-axis toward $+\infty$ (first SC) and $-\infty$ (second SC).  

We discretize the Hamiltonian in Eq. (\ref{HBdG}) using the finite difference method \cite{Smith2008} on a square lattice with lattice constant $a_x = a_y = 5~\text{nm}$ and solve numerically the corresponding BdG problem. The model parameters used in the calculation, $m^* = 0.026~m_e$, $\alpha_R = 200~\text{meV \AA}$, and $\Delta_o = 0.3~\text{meV}$, are consistent with InAs/Al structures in the strong SM-SC coupling limit. We note that the effective SM parameters may be renormalized in a strongly confined 2DEG, but incorporating such effects would require a more complicated modeling, e.g., using an 8-band Kane model \cite{Kane1957}. The discretized BdG Hamiltonian can be written in second quantized form as
\begin{equation}
    H = 
    \sum_{R,R^\prime} \sum_{i,j} \sum_{\mu,\nu}
    \psi^\dagger_{R,i,\mu} \mathcal{H}_{i\mu,j\nu}(R - R^\prime) \psi_{R^\prime,j,\nu}, \label{Hmtx}
\end{equation}
where $\psi^\dagger_{R,i,\mu}$ creates a fermion with spin/particle-hole index $\mu$ on site $i$ of the unit cell $R$, while $\mathcal{H}$ represents the corresponding Hamiltonian matrix. Note that the matrix elements are the same within every unit cell, as indicated by the $(R - R^\prime)$ dependence in Eq. (\ref{Hmtx}). Due to this periodicity, we can use Bloch's theorem by introducing the representation
\begin{equation}
    \psi_{R,i,\mu} = \frac{1}{\sqrt{N_x}} 
    \sum_q \widetilde{\psi}_{q,i,\mu} e^{-i q R} \label{psiR}
\end{equation}
into Eq. (\ref{Hmtx}). Note that in Eq. (\ref{psiR})  $q \in \left[-\pi,\pi\right]$ represents the crystal momentum, which is restricted to the first Brillouin zone, and $N_x$ is the number of unit cells along the length of the system. The BdG Hamiltonian takes the form
\begin{equation}
    H = 
    \sum_{q} \sum_{i,j} \sum_{\mu,\nu}
    \widetilde{\psi}^\dagger_{q,i,\mu} \widetilde{\mathcal{H}}_{i\mu,j\nu}(q) \widetilde{\psi}_{q,j,\nu}, \label{Hmtx2}
\end{equation}
where $\widetilde{\psi}_{q,i,\mu}^\dagger$ creates a fermion with spin/particle-hole index $\mu$ and crystal momentum $q$ on site $i$ of the unit cell. The Bloch Hamiltonian $\widetilde{\mathcal{H}}(q)$ is given by
\begin{equation}
    \widetilde{\mathcal{H}}_{i\mu,j\nu}(q) = 
    \sum_r \mathcal{H}_{i\mu,j\nu}(r) e^{iqr}. \label{H_Bloch}
\end{equation}
Note that the Bloch Hamiltonian has (anti-unitary) particle-hole symmetry, $\tau_x \widetilde{\mathcal{H}}^*(q) \tau_x = -\widetilde{\mathcal{H}}(-q)$, which is characteristic of all BdG Hamiltonians \cite{Chiu2016}, but time-reversal symmetry is broken due to the applied magnetic field. The system therefore belongs to class D and is characterized by a $\mathbb{Z}_2$ topological invariant $Q = \text{sign}\left(\text{Pf}[X(0)]\text{Pf}[X(\pi)]\right)$, where Pf indicates the Pfaffian and $iX(q)$ is the imaginary skew-symmetric Hamiltonian matrix in the Majorana basis \cite{Chiu2016}. The trivial SC phase corresponds to $Q = +1$, while the topological superconducting phase corresponds to $Q = -1$ and, in a finite system, gives rise to Majorana zero modes (MZMs) localized at the ends of the junction. Note that the phase boundaries separating the topological and trivial phases correspond to $\widetilde{\mathcal{H}}(q)$ having zero eigenvalues (i.e., gapless states) at $q=0$ or $q=\pi$. 
The key questions that we address within this theoretical framework are: 1) Given a set of structural parameters (e.g., $W_1$, $W_2$, $\ell$, and $L$), what is the corresponding topological phase diagram as function of the control parameters, $\mu$, $E_Z$, $V_J$, and $\phi$? 2) Within the topological region of parameter space, what is the size of the topological gap? Of course, a larger topological gap indicates a more robust topological phase and, implicitly, more robust MZMs.
Our main goal is to determine the impact of having a spatially-modulated junction width ($W_1\neq W_2$) on the extent and robustness of the topological phase.  

\subsection{Green's function formalism}

The hybrid system is perfectly well-defined by the Bloch Hamiltonian in Eqs. (\ref{Hmtx2}) and (\ref{H_Bloch}), but solving the corresponding quantum mechanics problem using a straightforward, brute force numerical procedure is at least challenging, or even impossible, if we assume that the proximitized regions are semi-infinite in the $y$-direction. 
We note that numerical calculations involving non-modulated structures (with $W_1=W_2$) do not face a similar challenge because the underlying problem is effectively one-dimensional and a scattering approach can be employed \cite{Pientka2017}. Alternatively, one can assume large, but finite, widths for the proximitized regions \cite{Hell2017a}, which implies solving a one-dimensional lattice problem for a system with $N_y$ sites, where $a (N_y+1)$ is the total width of the system. By contrast, using a similar approach for the modulated system involves a finite lattice with $N_L\times N_y$ sites, where $L = a N_L$ is the length of the unit cell, and the numerical problem becomes significantly more costly. To address this challenge, we use a self-energy approach within the Green's function formalism \cite{Odashima2016,Nolting2018} and integrate out the degrees of freedom associated with the large (possibly semi-infinite) proximitized regions. For quadratic Hamiltonians, the retarded Green's function is given by 
\begin{equation}
    G(\omega,q) = 
    \left(\omega - \widetilde{\mathcal{H}}(q) + i \eta\right)^{-1},
\end{equation}
where $\eta$ is an infinitesimally small positive energy that moves the poles of the retarded Green's function to the lower half of the complex plane. 
The reduced Green's function within the junction region is obtained by integrating out the proximitized regions. Specifically, we have
\begin{equation}
    G_J(\omega,q) = 
    \left(\omega - \widetilde{\mathcal{H}}_J(q)
    - \Sigma_{SC}(\omega,q)
    + i \eta\right)^{-1}, \label{GreensJunc}
\end{equation}
where the subscript $J$ indicates a quantity that is restricted to the junction region and $\Sigma_{SC}$ is the self-energy that captures the contribution of the two proximitized regions. The self-energy can be efficiently calculated numerically using the decimation method of Ref. \cite{Sancho1985}, which takes advantage of the fact that couplings between layers normal to the JJ interface are independent of the layer index.
Importantly, the topological index $Q$ can be calculated using the zero frequency  Green's function. Furthermore, the topological phase diagram can be efficiently calculated using the energy-independent effective Hamiltonian,
\begin{equation}
    \widetilde{\mathcal{H}}_{J}^{eff}(q) = 
    \widetilde{\mathcal{H}}_J(q)
    + \Sigma_{SC}(0,q).
\end{equation}
Indeed, the phase boundaries correspond to $G_J(0,q)$ having poles at $q=0$ or $q=\pi$, which is equivalent with $\widetilde{\mathcal{H}}_{J}^{eff}(q)$ having gapless modes at the corresponding values of $q$. Note that the problem is now numerically tractable since $\widetilde{\mathcal{H}}_{J}^{eff}(q)$ only contains the junction degrees of freedom. Also note that $\Sigma_{SC}(0,q)$ is Hermitian since there are no states within the SC gap for the isolated SC regions and, therefore, $\widetilde{\mathcal{H}}_{J}^{eff}(q)$ is also Hermitian. We also calculate the topological gap by finding the lowest-energy poles of Eq. (\ref{GreensJunc}) using the iterative method discussed in Appendix \ref{Iter}.

\begin{figure}[t]
\begin{center}
\includegraphics[width=0.48\textwidth]{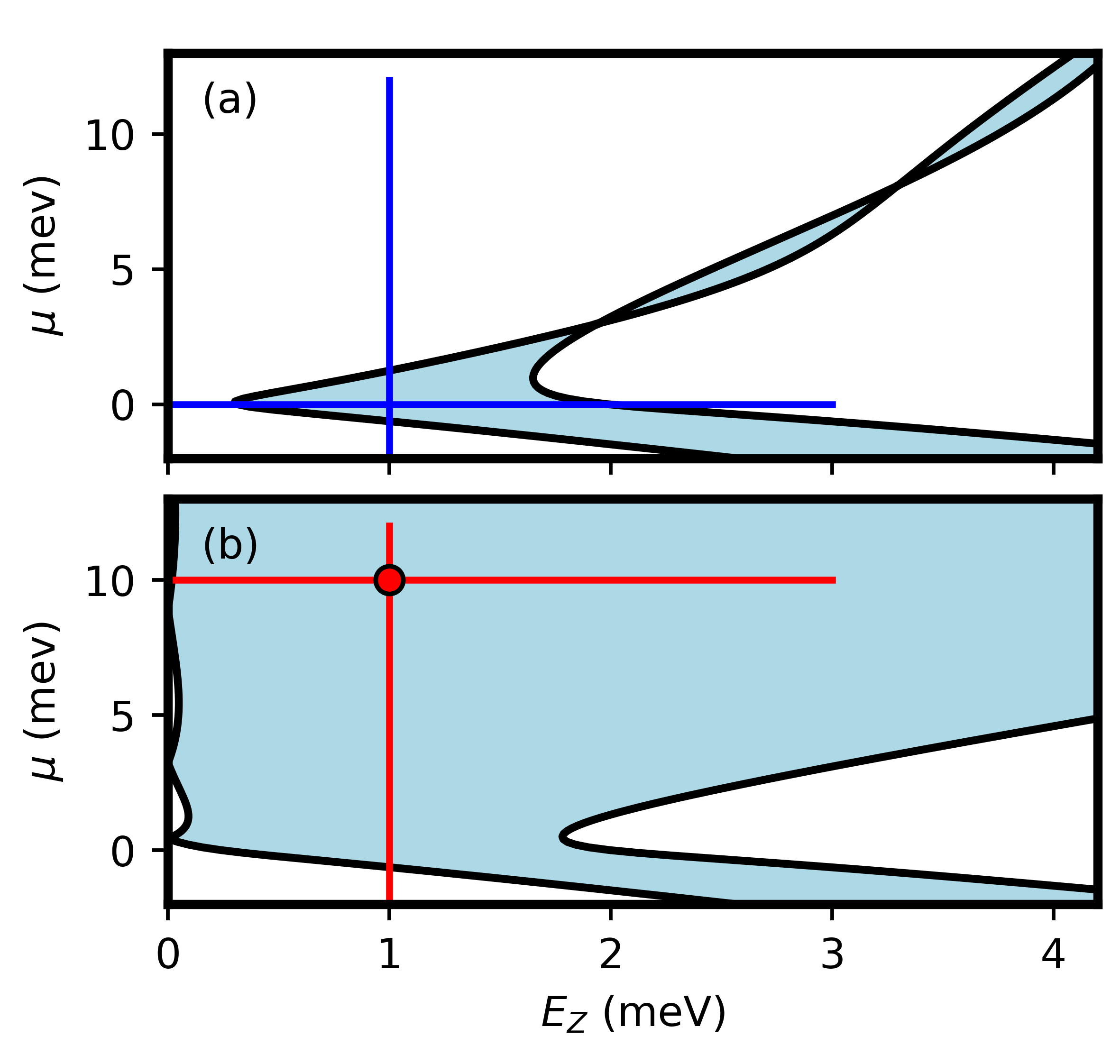}
\end{center}
\vspace{-0.6cm}
\caption{Phase diagram as a function of Zeeman energy and chemical potential for a uniform (i.e., non-modulated) system of width $W_1 = W_2 = 100~\text{nm}$. The superconducting phase bias is $\phi = 0, \pi$ in (a) and (b), respectively, and $V_J = 0$. Blue and red lines correspond to parameter values for which the topological gap is shown in Fig. \ref{FIG3}.}
\label{FIG2}
\vspace{-1mm}
\end{figure}

\section{Results} \label{Results}

\subsection{Uniform Josephson junctions} \label{NoMod}

\begin{figure}[t]
\begin{center}
\includegraphics[width=0.48\textwidth]{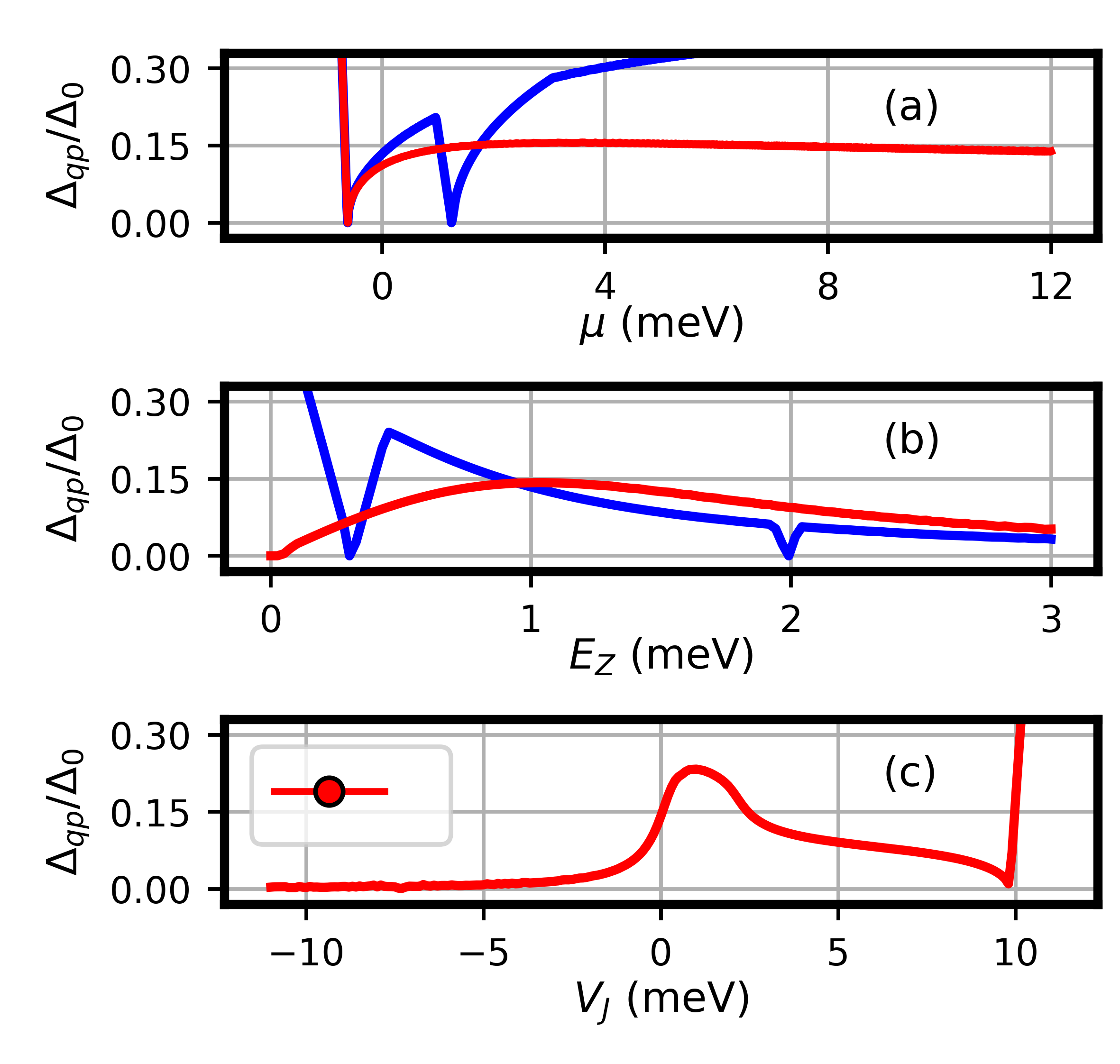}
\end{center}
\vspace{-0.6cm}
\caption{Quasi-particle gap along representative cuts marked by blue and red lines in  Fig. \ref{FIG2} as functions of (a) chemical potential, $\mu$, and (b) Zeeman splitting, $E_z$. Blue lines correspond to $\phi = 0$, while red lines correspond to a system with superconducting phase difference $\phi = \pi$. Panel (c) shows the dependence of the quasiparticle gap on the applied junction potential, $V_J$, for a system with parameters corresponding to the red dot in Fig. \ref{FIG2} (b). Note that the maximum topological gap for the uniform (non-modulated) system is $\Delta_{\text{top}} \approx 0.23 \Delta_o = 0.07~\text{meV}$.}
\label{FIG3}
\vspace{-1mm}
\end{figure}

We first consider a uniform, non-modulated system, which provides reference results for evaluating the modulated structures. This case also illustrates some of the potential concerns about (uniform) Josephson junction structures, in particular regarding the size of the topological gap that can be realized in this type of system.
An example of topological phase diagram as function of the Zeeman splitting ($E_Z$) and chemical potential ($\mu$) for a system of junction width $W_1 = W_2 = 100~\text{nm}$ and superconducting phase difference (a) $\phi = 0$ and (b) $\phi = \pi$ is shown in Fig. \ref{FIG2}. 
Note that the junction potential is set to zero, $V_J = 0$. The phase diagrams are consistent with previous studies \cite{Hell2017a,Pientka2017}. We note that the experimentally-relevant regime corresponds to relatively low values of the Zeeman splitting (e.g., $E_Z \lesssim 2~$meV), since applying large magnetic fields is detrimental to superconductivity inside the parent SC (i.e., the Al films that proximitize the 2DEG) and, implicitly, to the size of the topological gap. As expected \cite{Hell2017a,Pientka2017}, in the presence of a phase difference $\phi=\pi$ practically the entire the low-field region of the phase diagram with $\mu \gtrapprox 0$ is covered by the topological superconducting phase. 
By contrast,  the system with $\phi = 0$ (i.e., no phase difference) is characterized by a single, relatively narrow topological lobe around $\mu\approx 0$. Additional small topological regions occur at higher $\mu$ and $E_Z$  values, but, because they are (mostly) outside the low-field regime, these regions are expected to have rather limited experimental significance. 
This behavior is due to the fact that the relevant low-energy states leak further into the proximitized regions as $\mu$ increases and, therefore, require larger values of $E_Z$ (which is nonzero only inside the junction region) to acquire the effective Zeeman splitting consistent with the emergence of topological superconductivity. 

\begin{figure}[t]
\begin{center}
\includegraphics[width=0.48\textwidth]{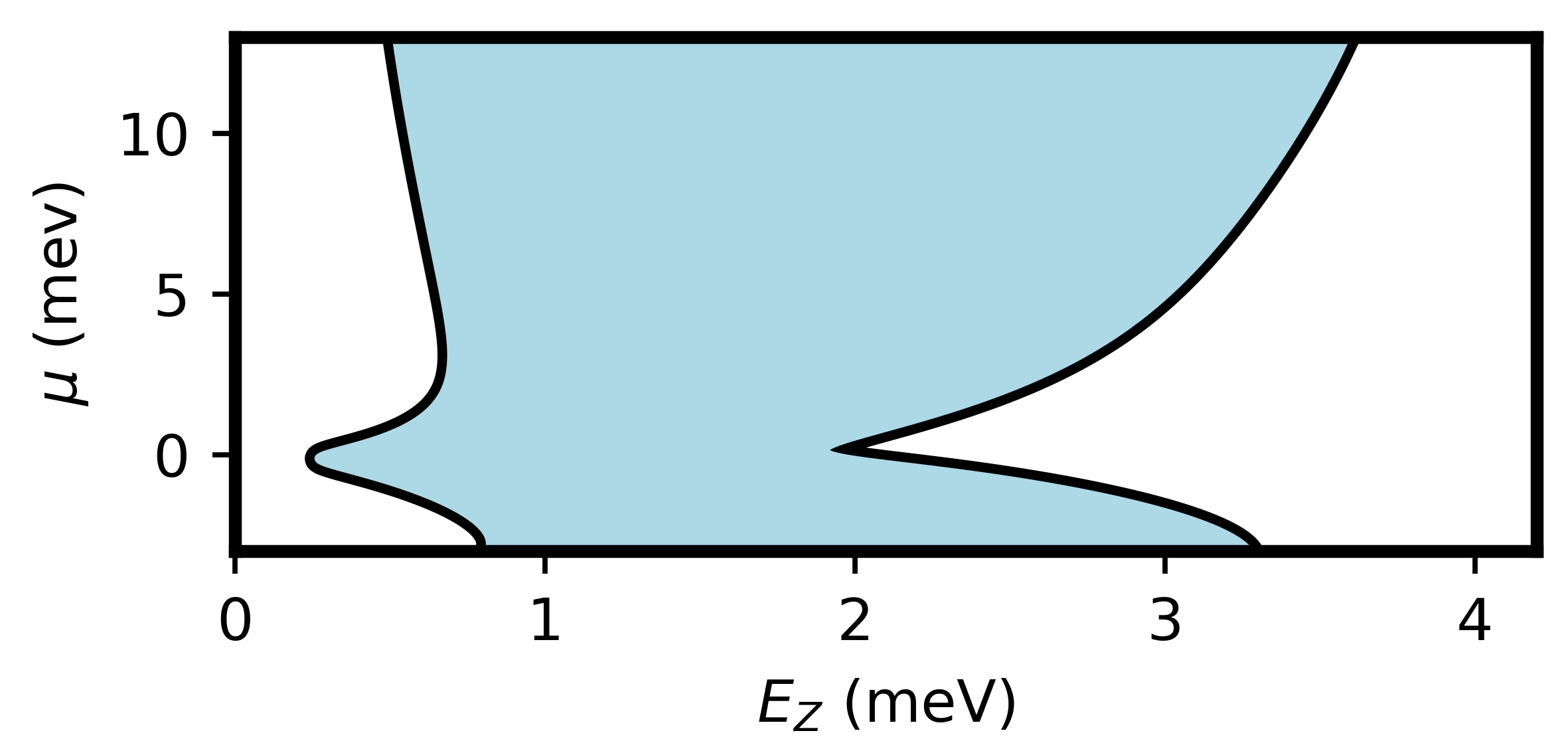}
\end{center}
\vspace{-3mm}
\caption{Topological phase diagram for a uniform system of width $W_1 = W_2 = 100~\text{nm}$ with no superconducting phase difference ($\phi=0$) and optimal junction potential, $V_J=V_J^*(\mu)$ (see main text). Note that the lowest critical Zeeman field in a system with finite chemical potential is always larger than the minimal value $E_Z = \Delta_o=0.3~$meV, which corresponds to $\mu=0$.}
\label{FIG4}
\vspace{-3mm}
\end{figure}

The quasi-particle gap  $\Delta_{qp}$ (defined as the lowest positive eigenenergy of the bulk spectrum) corresponding to the cuts marked by blue lines in Fig. \ref{FIG2} (a) are shown as blue lines in Fig. \ref{FIG3} (a) and (b). Note that $\Delta_{qp}$ is the topological gap, $\Delta_{\text{top}} = \Delta_{qp}$, when the system is in the topological phase. The maximum topological gap along these representative cuts for the system with no phase difference ($\phi = 0$) is $\Delta_{\text{top}} \approx 0.23 \Delta_o = 0.07~\text{meV}$.
The quasi-particle gap corresponding to the red lines in Fig. \ref{FIG2} (b), i.e., for a system with phase difference $\phi=\pi$, are shown as red lines in Fig. \ref{FIG3} (a) and (b). 
Note that, as a function of the applied Zeeman field, the topological gap has a maximum $\Delta_{\text{top}} \approx 0.15 \Delta_o = 0.045~\text{meV}$ at $E_Z \approx 1~\text{meV}$ [see Fig. \ref{FIG3} (b)]. The dependence of this maximum value on $\mu$ is rather weak, as shown  in Fig. \ref{FIG3} (a).  
The effect of a nonzero junction potential $V_J$ on the topological gap is shown in 
Fig. \ref{FIG3} (c) for parameters corresponding to the red dot in Fig. \ref{FIG2} (b). Note that for $V_J < 0$ the topological gap quickly decreases toward zero. This is due to the formation of bound states localized almost entirely within the junction region, which are characterized by a small induced gap, as discussed in Appendix \ref{QEM}. The maximum of the topological gap is obtained for $V_J \approx 1~$meV and has a value $\Delta_{\text{top}} \approx 0.23 \Delta_o= 0.07~\text{meV}$ comparable to the maximum gap of the system with no phase difference ($\phi=0$). Upon further increasing $V_J$  the system becomes non-topological near $V_J \approx 10~\text{meV}$, when the junction region becomes depleted (i.e. $V_J > \mu$).

An important question is whether the low-field topological lobe characterizing the system with $\phi=0$ can be accessed by tuning $V_J$ when the system has a finite (possibly large) chemical potential. To address this question, we determine the value $V_J^*(\mu)$ of the junction potential that minimizes the critical Zeeman field for a given value of the chemical potential. 
We find that $V_J^*(\mu) \lessapprox \mu$, i.e., the optimal $V_J$ is slightly smaller than the value of the chemical potential. The topological phase diagram as a function of $E_Z$ and $\mu$ for a system with optimal junction potential $V_J=V_J^*(\mu)$ is shown in Fig. \ref{FIG4}. Note that the phase boundary shifts from a minimum $E_Z = \Delta_o=0.3~$meV at $\mu=0$ to larger values of the critical Zeeman splitting when $\mu\neq 0$, which can make accessing the topological phase more difficult. We emphasize that the experimentally-relevant situation corresponds to $\mu>0$, which implies that the  minimal critical field (obtained by tuning $V_J$) in a system with no phase difference (i.e., $\phi=0$) is always larger than (but comparable to) $\Delta_o$. This analysis shows that, as far as the accessibility of the topological phase is concerned, there is no fundamental advantage of being able to apply a nonzero phase difference if the regime $E_Z\sim 1~$meV is experimentally accessible. In other words, there is no major difference between accessing the topological quantum states in Fig. \ref{FIG3}(b) by tuning the phase difference and accessing the topological quantum states in Fig. \ref{FIG4} by tuning the junction potential. If, on the other hand, $E_Z\sim 1~$meV is outside of the low-field regime, having $\phi =\pi$ may be a significant advantage. However, this comes with the heavy price of a small topological gap (see Fig. \ref{FIG3}).

\begin{figure}[t]
\begin{center}
\includegraphics[width=0.48\textwidth]{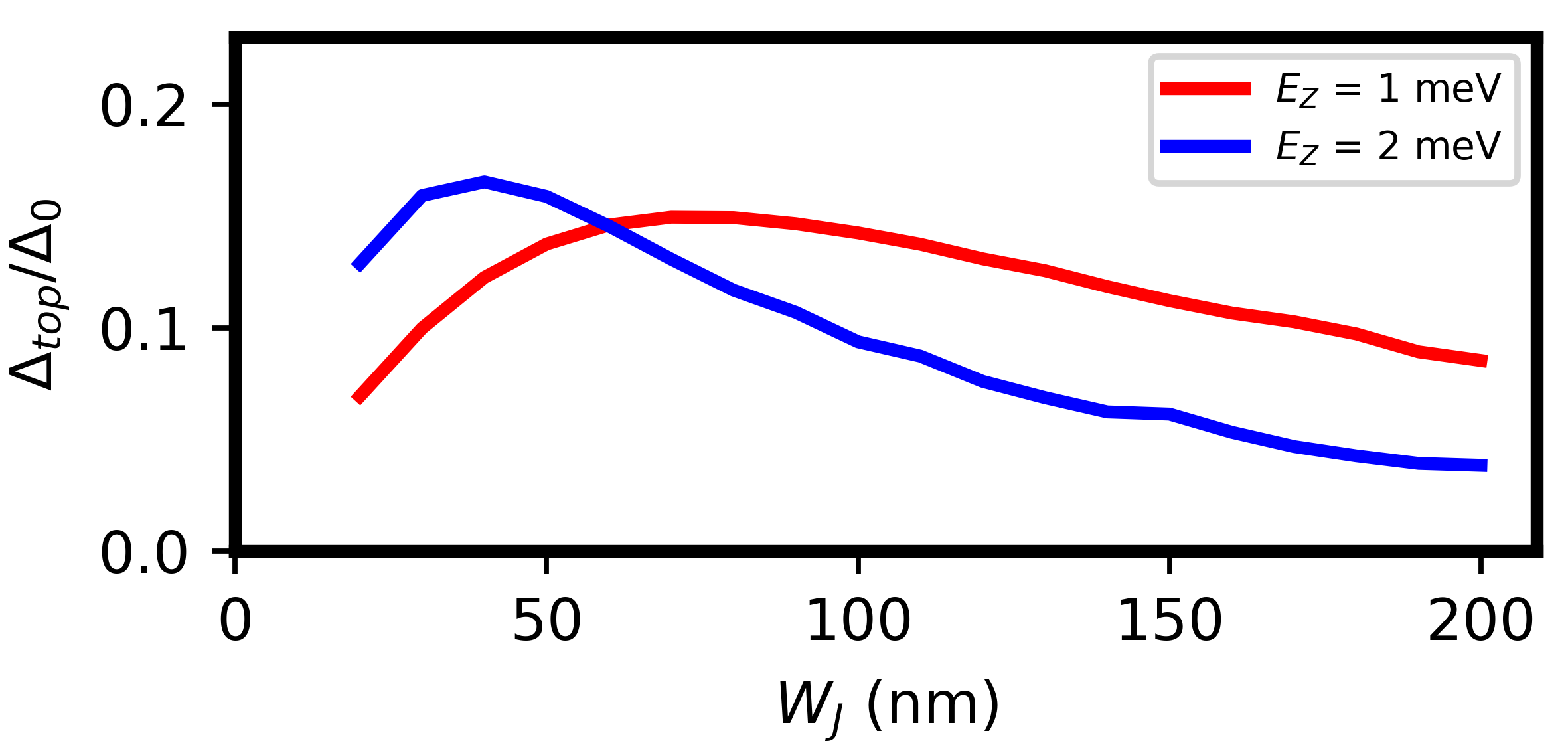}
\end{center}
\vspace{-3mm}
\caption{Dependence of the topological gap on the junction width $W$ for a uniform system with chemical potential $\mu = 10~\text{meV}$, phase difference $\phi = \pi$, and two values of the Zeeman field: $E_Z = 1~$meV (red) and $E_Z=2~$meV (blue). Note that the dependence is non-monotonic. Reducing the width of the junction allows larger values of the maximum topological gap, but this requires larger Zeeman fields that may be outside the (experimentally-relevant) low-field regime.}
\label{FIG5}
\vspace{-3mm}
\end{figure}

While applying a phase difference $\phi=\pi$ can solve potential problems regarding the accessibility of the topological phase due to finite critical values of $E_Z$, there remains the key issue of the relatively small topological gap. Why is the (maximum) topological gap only a small fraction of $\Delta_o$ in non-modulated systems? As pointed out in Ref. \cite{Pientka2017}, the topological gap is expected to be large if $W \lesssim \sqrt{\hbar^2/(m^* \Delta_o)}$;  otherwise, it is inversely proportional to the square of the junction width,  $\Delta_{\text{top}} \propto W^{-2}$. For the values of the effective mass and induced gap used in this work, the large gap condition yield $W \lesssim 100~\text{nm}$, which suggests that a larger topological gap could be obtained by using narrower junctions.
This motivates us to calculate the dependence of the topological gap on the junction width. Fig. \ref{FIG5} shows $\Delta_{\text{top}}$ as a function of the junction width for a uniform system with $\mu=10~$meV, $\phi=\pi$ and two different values of $E_Z$. Note that the topological gap decreases with increasing the junction width for large-enough $W$ values, as expected based on the asymptotic $\Delta_{\text{top}} \propto W^{-2}$ behavior. Perhaps more surprising is the suppression of the gap in the narrow junction limit, $W\rightarrow 0$. This occurs because the effective Zeeman energy associated with the relevant low-energy states is proportional to the spectral weight of the states within the junction, which is reduced as $W$ decreases. For small $W$ values, the topological gap will reach its maximum  at a higher  $E_Z$, which may be outside of the (experimentally-relevant) low-field regime. 
The maximum gap corresponding to a given junction width $W$ occurs at a value of the Zeeman splitting that increases with decreasing $W$. Consequently, 
reducing $W$ cannot be a practical solution to the problem of small topological gaps in uniform systems. Finally, we note that an enhancement of the topological gap could be obtained by significantly increasing the spin-orbit coupling strength \cite{Pientka2017}. However, we find that reaching values of the maximum topological gap above $\Delta_{\text{top}} \approx 0.3 \Delta_o$ would be extremely difficult within any realistic parameter regime. In the next section we show that a practical solution to ``artificially''  enhancing the effective spin-orbit coupling involves a periodic potential created by modulating the width of the Josephson junction.
 
To summarize this section, we point out that the most appealing feature of the Josephson junction proposal for realizing topological superconductivity and MZMs -- the extensive low-field topological phase emerging in the presence of a phase difference $\phi=\pi$ [see Fig. \ref{FIG2} (b)] -- is offset by serious limitations regarding the size of the topological gap. For the realistic parameters used in our calculation, the maximum topological gap is $\Delta_{\text{top}} \approx 0.23 \Delta_o= 0.07~\text{meV}$. Moreover, generating this gap requires not only a finite Zeeman field, $E_Z\approx 1~$meV, but also tuning the Junction potential $V_J$ (see Fig. \ref{FIG3}). However, if the regime $E_Z\approx 1~$meV is accessible and $V_J$ can be tuned, one can realize similar values of the topological gap in a system with no phase difference, by simply tuning the junction potential (see Figs. \ref{FIG3} and \ref{FIG4}). In both cases the major problem is the relatively small topological gap and, consequently, the fragility of the topological phase and of the emerging MZMs. 

\subsection{Modulated Josephson junctions}

In this section we present the numerical results for the  proposed modulated Josephson junction  structure and show (see Sec. \ref{phi0}) that, in the absence of a superconducting phase difference (i.e., for $\phi=0$) the system operated in the quantum well regime ($V_J <0$)  (i) supports a low-field topological phase that covers a significant area of the phase diagram and (ii) is characterized by an enhanced  topological gap that represents a substantial fraction of the induced gap $\Delta_o$. For completeness, we also consider the case $\phi=\pi$ and show (see Sec. \ref{phiPi}) that having a superconducting phase difference $\phi=\pi$ provides no practical advantage in a system with modulated junction width. We emphasize that the calculations presented in this section should be considered as  ``proof-of-concept'' examples . Note that we do not explicitly address the issue of optimizing the geometric parameters; this optimization task should be addressed in synergy with the materials growth and structure engineering efforts and should target specific materials and hybrid structures. 

\begin{figure}[t]
\begin{center}
\includegraphics[width=0.48\textwidth]{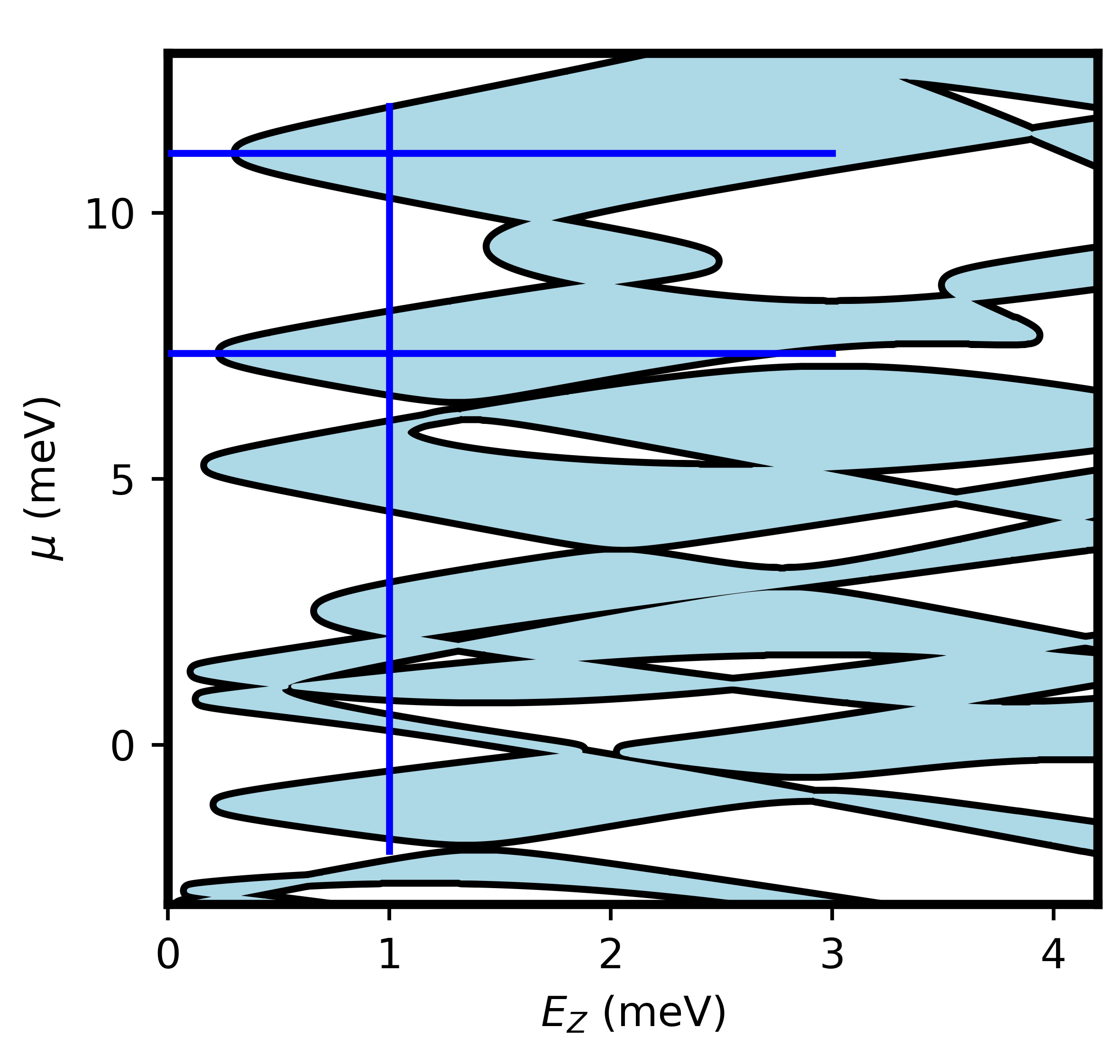}
\end{center}
\vspace{-3mm}
\caption{Phase diagram of a modulated structure with no superconducting phase difference ($\phi=0$) as a function of Zeeman energy, $E_Z$, and chemical potential, $\mu$. The geometric parameters are $W_1 = 100~\text{nm}$, $W_2 = 20~\text{nm}$, $L = 60~\text{nm}$, $\ell = 20~\text{nm}$, and $w = 40~\text{nm}$ (see Fig. \ref{FIG1}). The junction potential is negative, $V_J = -40~\text{meV}$,  which generates several bound states within the junction region. As compared to the  phase diagram for the uniform system [see in Fig. \ref{FIG2} (a)], the low-field topological region is dramatically expanded; there exist several topological lobes at low $E_Z$ values, each associated with a different folded subband. The quasi-particle  gap corresponding the parameter cuts  indicated by the blue lines  are shown in Figs. \ref{FIG7} and \ref{FIG8}.}
\label{FIG6}
\vspace{-3mm}
\end{figure}

\subsubsection{Modulated Josephson junctions with no phase difference ($\phi=0$)} \label{phi0}

Let us consider a two-dimensional semiconductor-superconductor hybrid system with no superconducting phase difference, $\phi = 0$, in the presence of a spatial  modulation characterized by  the geometric parameters  $W_1 = 100~\text{nm}$, $W_2 = 20~\text{nm}$, $L = 60~\text{nm}$, and $\ell = 20~\text{nm}$. We assume that the system has reflection symmetry about the $x$-axis. 
A negative junction potential $V_J = -40~\text{meV}$ puts the system into the quantum well regime. In a uniform structure, the negative junction potential would generate a trivial topological phase, if the phase difference is zero, or an essentially  gapless topological phase at $\phi =\pi$,  as shown numerically  in Fig. \ref{FIG3} (c) (see also the discussion in Appendix  \ref{QEM}). 
However, modulating the junction width significantly enhances the quasiparticle gap. The corresponding phase diagram, as a function of $E_Z$ and $\mu$, is shown in Fig. \ref{FIG6}. First, note that the phase diagram is quite complicated, especially at larger values of $E_Z$ (i.e., $E_Z > 2~$meV). This behavior is  due to the presence of many  folded subbands, which  often cross giving rise to a rather atypical phase diagram. Nonetheless, the crucial point is that the low-field  topological region of the phase diagram is dramatically enlarged as compared to the corresponding phase diagram of the non-modulated structure  [see Fig. \ref{FIG2} (a)]. We emphasize that, unlike the uniform structure, in this modulated system the topological phase  exists at low Zeeman energies, $E_Z \sim \Delta_o$, even for relatively large $\mu$ values. This substantial enhancement of the low-field topological region represents the first significant advantage of the modulated structures.

\begin{figure}[t]
\begin{center}
\includegraphics[width=0.48\textwidth]{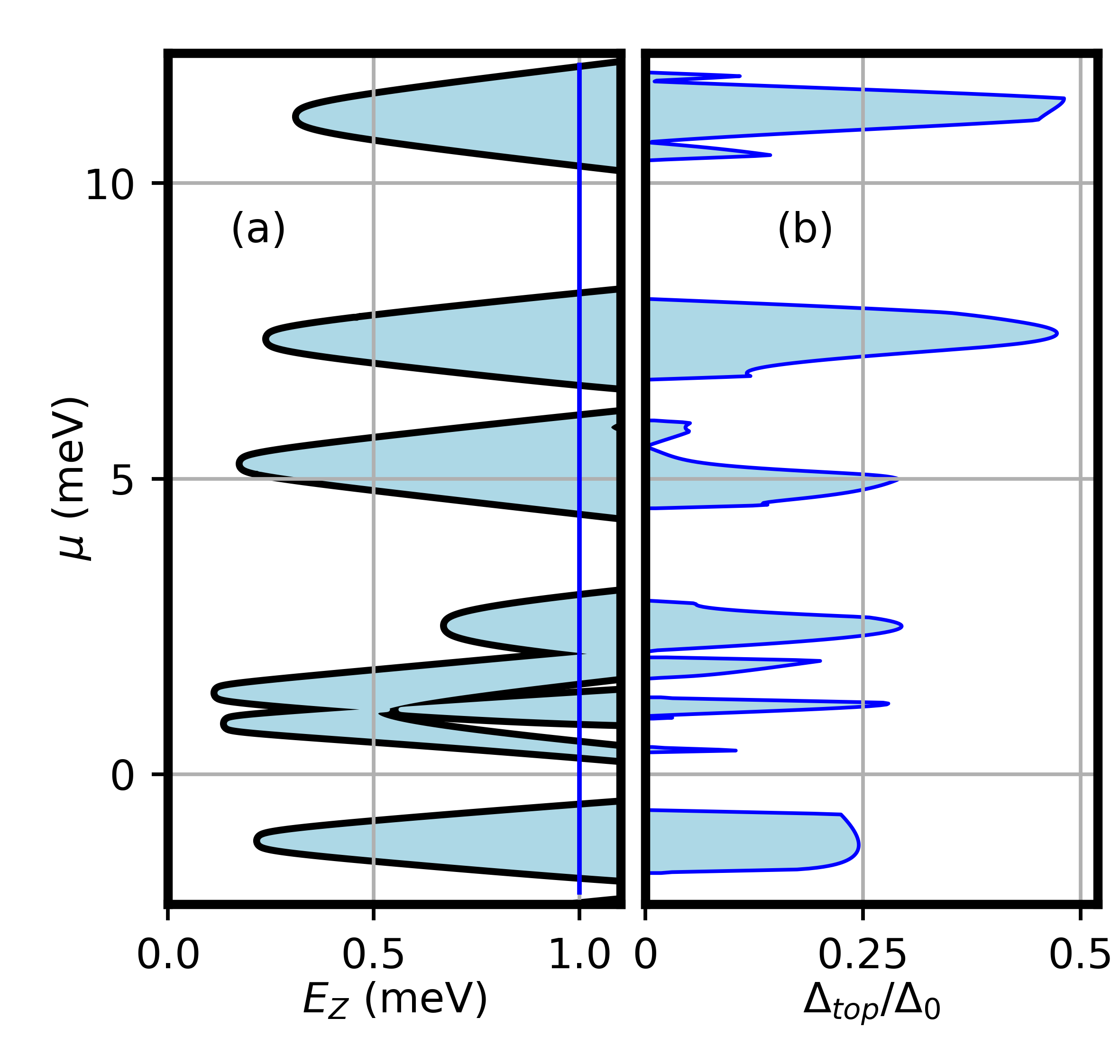}
\end{center}
\vspace{-3mm}
\caption{(a) Low-field phase diagram corresponding to the region $E_Z < 1.1~$meV in Fig. \ref{FIG6}. Note the presence of several topological lobes (shaded) with $E_{Z,crit} \sim \Delta_o$. (b) Topological gap as a function of the chemical potential $\mu$  for fixed $E_Z = 1~\text{meV}$ [blue line in (a)]. The quasiparticle gap corresponding to the topologically-trivial phase is not shown. Note that the maximum topological gap approaches $0.5 \Delta_o$, which represents a significant increase as compared to the non-modulated structure (see blue lines in Fig. \ref{FIG3}).}
\label{FIG7}
\vspace{-3mm}
\end{figure}

To asses the robustness of the topological phase, we  calculate the topological gap along representative cuts through the parameter space, wich are shown as blue lines in Fig. \ref{FIG6}. The results for the vertical cut (i.e., the dependence on the chemical potential $\mu$ for a  fixed value of the Zeeman splitting $E_Z = 1~\text{meV}$) are shown in Fig. \ref{FIG7}. For convenience, the low-field portion of the phase diagram shown in  Fig. \ref{FIG6} is reproduced in panel (a), while panel (b) shows the topological gap along the vertical cut marked by the blue line. For simplicity, we do not plot the quasiparticle gap corresponding to the topologically-trivial superconducting phase.
Note that the low-field  phase diagram, especially for $\mu > 4~\text{meV}$, is qualitatively similar to  a ``conventional'' Majorana phase diagram for a multi-subband hybrid system \cite{Stanescu2013}. In our case, the subbands responsible for the emergence of the topological lobes  are actually ``minibands'' resulting from  the folding of bound state bands (i.e., the bands associated with states localized near the junction region) into the first Brillouin zone associated with the periodic structure. 
The topological gap  for a  fixed  value of the Zeeman field, $E_Z = 1~\text{meV}$, is shown in Fig. \ref{FIG7} (b). Note that  the topological gap corresponding to  the top two lobes is quite large, having peak values  $\Delta_{\text{top}} \approx 0.47 \Delta_o$. This is a significant increase  (by a factor of two) as compared to the uniform system , which has a maximum topological gap of $\Delta_{\text{top}} \approx 0.23 \Delta_o$. We note that this enhancement was not optimized with respect to the geometric parameters characterizing the modulated junction, or with respect to the applied junction potential $V_J$, and is obtained at a value of the applied Zeeman field comparable to the (optimal) low-field values associated with the uniform system. 

\begin{figure}[t]
\begin{center}
\includegraphics[width=0.48\textwidth]{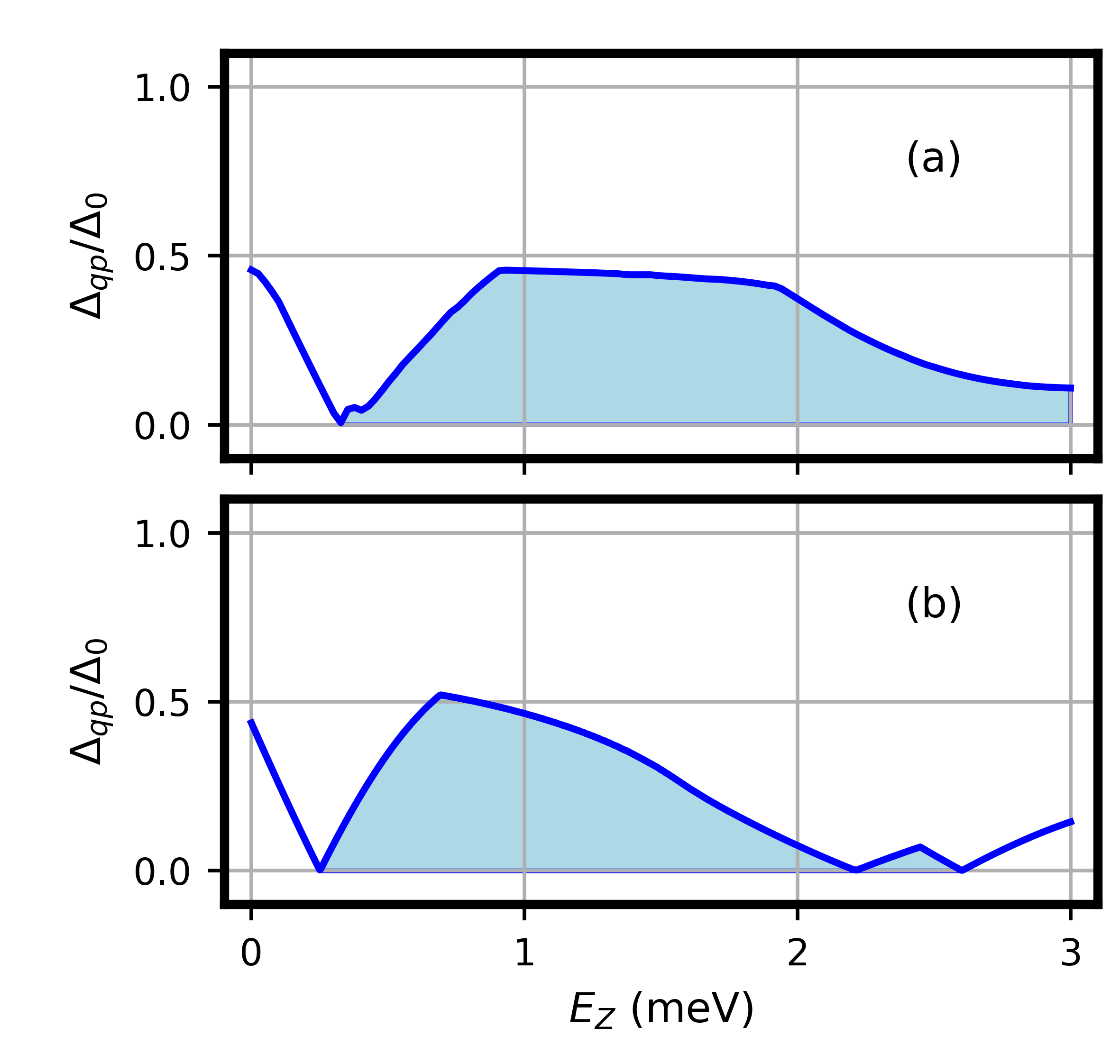}
\end{center}
\vspace{-3mm}
\caption{Quasi-particle gap as a function of Zeeman energy corresponding to the horizontal lines in Fig. \ref{FIG6} with: (a) $\mu = 11.1~$meV and (b) $\mu = 7.4~$meV. Shading  indicates the presence of a topological superconducting phase, i.e. $\Delta_{qp} = \Delta_{\text{top}}$. Note that the topological gap remains near its maximum value, $\Delta_{\text{top}} \approx 0.5 \Delta_o$, over a significant range of $E_Z$.}
\label{FIG8}
\vspace{-3mm}
\end{figure}

The enhanced topological gap  exists over a larger range of Zeeman field, as shown  in Fig. \ref{FIG8}. Furthermore, the maximum value corresponding to the parameter range used in this calculation slightly exceeds $0.5\Delta_o$ and is obtained at a relatively low field, $E_Z \approx 2.2\Delta_o=0.66~$meV. These results clearly show that modulating the junction width can lead to a substantial enhancement of the topological gap.
The key physical mechanism responsible for the enhancement of the topological gap is associated with the increase of the effective spin-orbit coupling in the presence of a periodic potential \cite{Woods2020b}. More specifically,  the high-order minibands (i.e., the minibands formed from subbands that have folded several times) are characterized by a substantially enhanced effective spin-orbit coupling \cite{Woods2020b}, which, in turn, produces a large topological gap. 
Note that the lower-order minibands, which are responsible for the topological regions at smaller $\mu$ values, are characterized by a weaker enhancement of the effective spin-orbit coupling and, consequently, a weaker enhancement of the topological gap. On the other hand,  further increasing the chemical potential ($\mu>12~$meV) leads to the emergence of topological lobes generated by even higher-order minibands, with extremely large effective spin-orbit coupling, which can push the maximum topological gap closer to $\Delta_o$.

\begin{figure}[t]
\begin{center}
\includegraphics[width=0.48\textwidth]{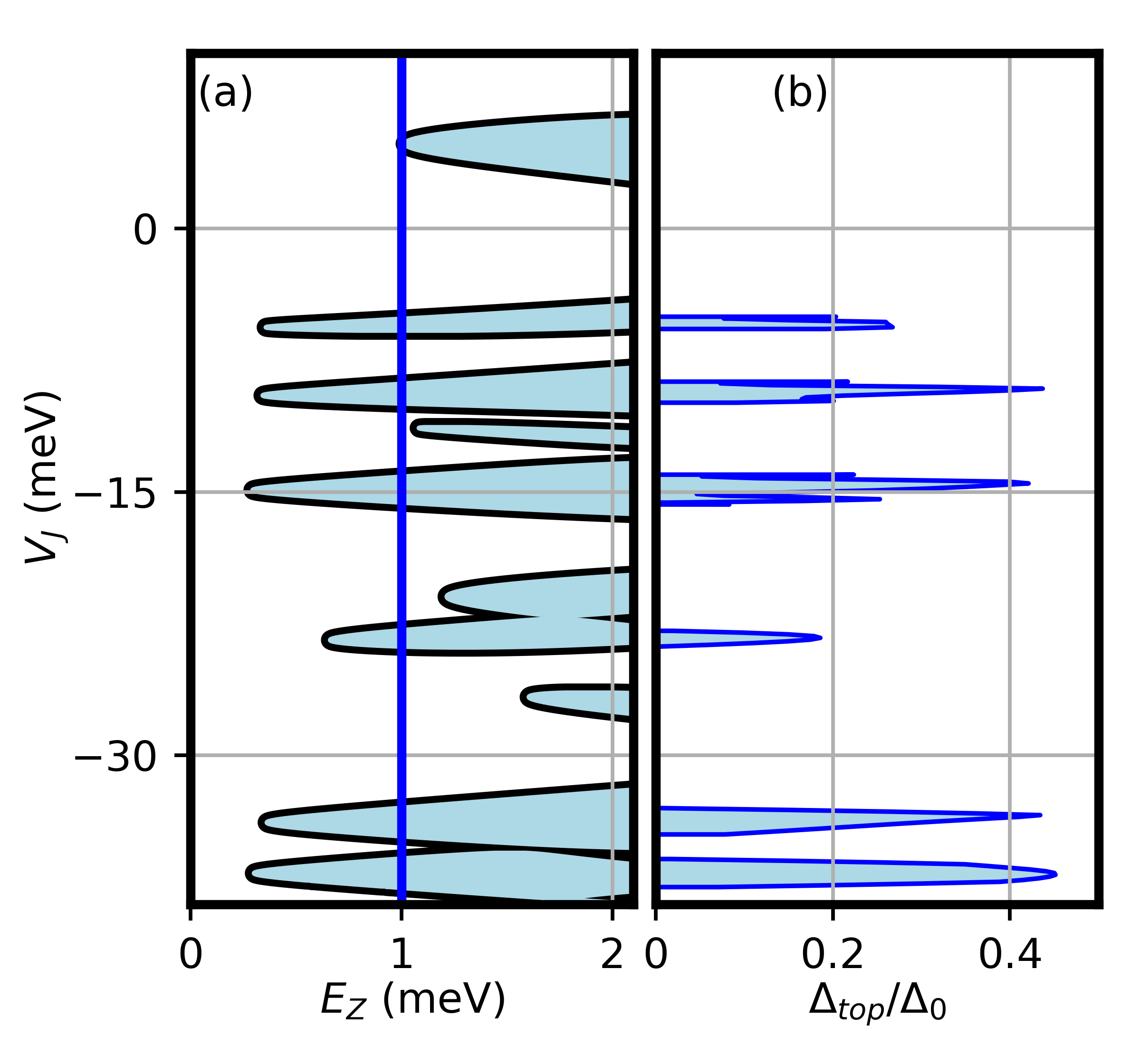}
\end{center}
\vspace{-3mm}
\caption{(a) Phase diagram of a modulated structure with the same parameters as in Fig. \ref{FIG6} and $\mu = 10~$meV as function of the Zeeman field, $E_Z$, and the applied junction potential, $V_J$. (b) Topological gap along the cut marked by a blue line in (a), wich corresponds to a Zeeman field $E_Z=1~$meV. The trivial quasiparticle gap is not shown.}
\label{FIG18}
\vspace{-3mm}
\end{figure}

The results presented in Figs. \ref{FIG6}, \ref{FIG7}, and \ref{FIG8} illustrate the main benefits of engineering hybrid structures with periodically modulated junction width: the emergence of multiple low-field topological lobes in the absence of a superconducting phase difference,  $\phi = 0$, and the  enhancement of the topological gap. 
However, for a given structure the chemical potential $\mu$ is not an experimentally  tunable parameter. In fact, $\mu$ is essentially determined by details of the 2DEG-SC heterostructure, such as materials properties, band-bending effects at the SM-SC interface \cite{Woods2018,Winkler2019}, and the strength of the effective SM-SC coupling \cite{Stanescu2017a,Reeg2018a}. On the other hand, the junction potential $V_J$ is readily tunable using a top gate and, therefore, represents an experimentally- relevant control parameter, along with the Zeeman field $E_Z$.
The low-field topological lobes generated by the (high-order) minibands and characterized by large values of the topological gap should be accessed by tuning $V_J$, rather than the chemical potential. 

An example of phase diagram as function of Zeeman field, $E_Z$, and applied junction potential, $V_J$, for a system with the same structure parameters as in Fig. \ref{FIG6} and a (fixed) chemical potential  $\mu = 10~\text{meV}$ is shown in Fig. \ref{FIG18} (a). The phase diagram is characterized by several topological regions with low values of the critical Zeeman field, $E_Z \sim \Delta_o$. As $V_J$ is tuned toward more negative values in the presence of a Zeeman field of order $E_Z\approx 1~$meV, we  sweep through several topological lobes characterized by large values of the topological gap, $\Delta_{\text{top}} \approx 0.4 - 0.45\Delta_o$ [see Fig. \ref{FIG18}(b)]. This represents a significant enhancement of topological gap as compared to the non-modulated system [see Fig. \ref{FIG2} (a)]. 
We also note the presence of a topological region for positive $V_J$ values (i.e., in the potential barrier regime) near $V_J \approx 5~\text{meV}$.  However, the correponding critical Zeeman energy is $E_Z \approx 1~\text{meV}$, significantly larger than the lowest critical Zeeman energies in the quantum well regime ($V_J <0$). This behavior is due to a stronger localization of the low-energy states within the junction region in the quantum well regime as compared to the potential barrier case. In turn, since the Zeeman splitting is assumed to be small in the proximitized regions, this results in a larger effective Zeeman energy associated with a given value of $E_Z$ for the system in the quantum well regime ($V_J < 0$). 
The results shown in Fig. \ref{FIG18} demonstrate that the enhanced topological regions generated  by minibands emerging  in the presence of a modulation-induced periodic potential can be conveniently accessed by tuning the junction potential $V_J$. We emphasize that tuning $V_J$ using a potential gate can be done much more efficiently than tuning the effective potential of a hybrid nanowire structure, because of minimal screening by the superconductor. Indeed, the ``active region'' in a nanowire is relatively close to the superconductor, which drastically limits the effectiveness of a potential gate. By contrast, the junction region is free from this limitation, which enables tuning $V_J$ within a large potential window and, consequently, exploring large regions of the parameter space. This is useful not only for optimizing the topological superconducting phase (by maximizing the topological gap), but also for investigating topological quantum phase transitions. Of course, the presence of disorder can seriously limit or completely destroy this physics.

\subsubsection{Modulated Josephson junctions with phase difference $\phi=\pi$} \label{phiPi}

\begin{figure}[t]
\begin{center}
\includegraphics[width=0.48\textwidth]{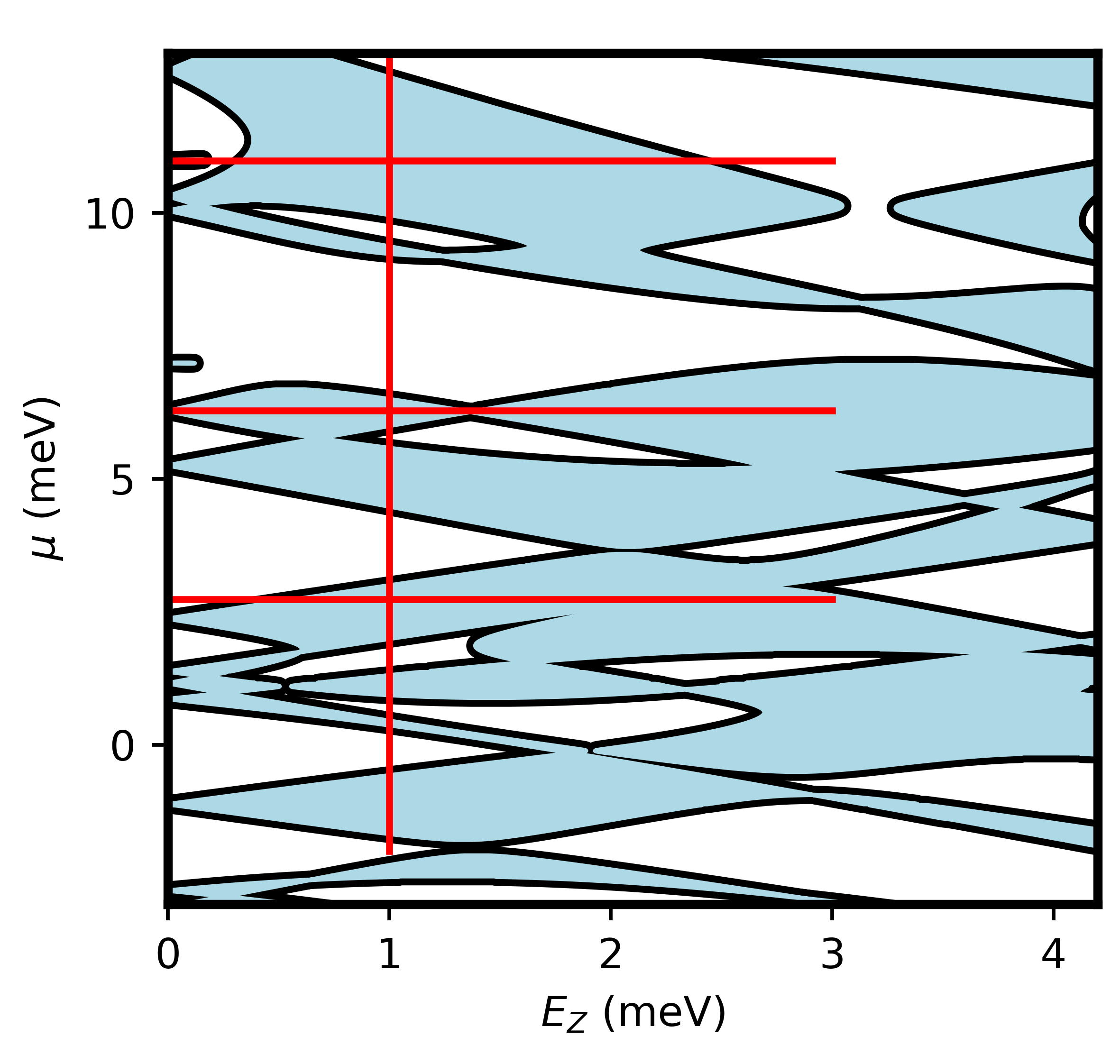}
\end{center}
\vspace{-3mm}
\caption{Phase diagram of a modulated structure with superconducting phase difference $\phi=\pi$ as a function of Zeeman energy, $E_Z$, and chemical potential, $\mu$. The geometric parameters are the same as in Fig. \ref{FIG6} and the junction potential is negative, $V_J = -40~\text{meV}$. The quasiparticle gap calculated along the cuts marked by red lines is shown in Figs. \ref{FIG10} (b) and \ref{FIG11}.}
\label{FIG9}
\vspace{-3mm}
\end{figure}

Having demonstrated the enhancement of the topological gap in modulated structures with no superconducting phase difference ($\phi =0$), the natural question regards the fate of topological superconductivity in the presence of a  phase difference $\phi = \pi$. We note that, in general, modulating the junction with of a system operated in the potential barrier regime ($V_J \geq 0$) generates no advantage with respect to the uniform structure (see the discussion in Appendix \ref{QEM}) and, therefore, we focus on the quantum well regime.
Fig. \ref{FIG9} shows an example of topological phase diagram as a function of $E_Z$ and $\mu$ for a system with the same parameters as in Fig. \ref{FIG6} and phase difference  $\phi = \pi$. Similarly to the non-modulated system in Fig. \ref{FIG2}, the phase boundary shifts  toward lower values of the Zeeman field, reaching $E_Z = 0$ at certain values of the chemical potential. 
The phase diagram exhibits an alternation of topological and trivial phases as the chemical potential is varied in the presence of a finite Zeeman field. Note that the ranges of $\mu$ where the phase boundary approaches $E_Z = 0$ correspond to the low $E_Z$ topological lobes  in Fig. \ref{FIG6}, suggesting that they are associated with the presence of modulation-induced minibands.

\begin{figure}[t]
\begin{center}
\includegraphics[width=0.48\textwidth]{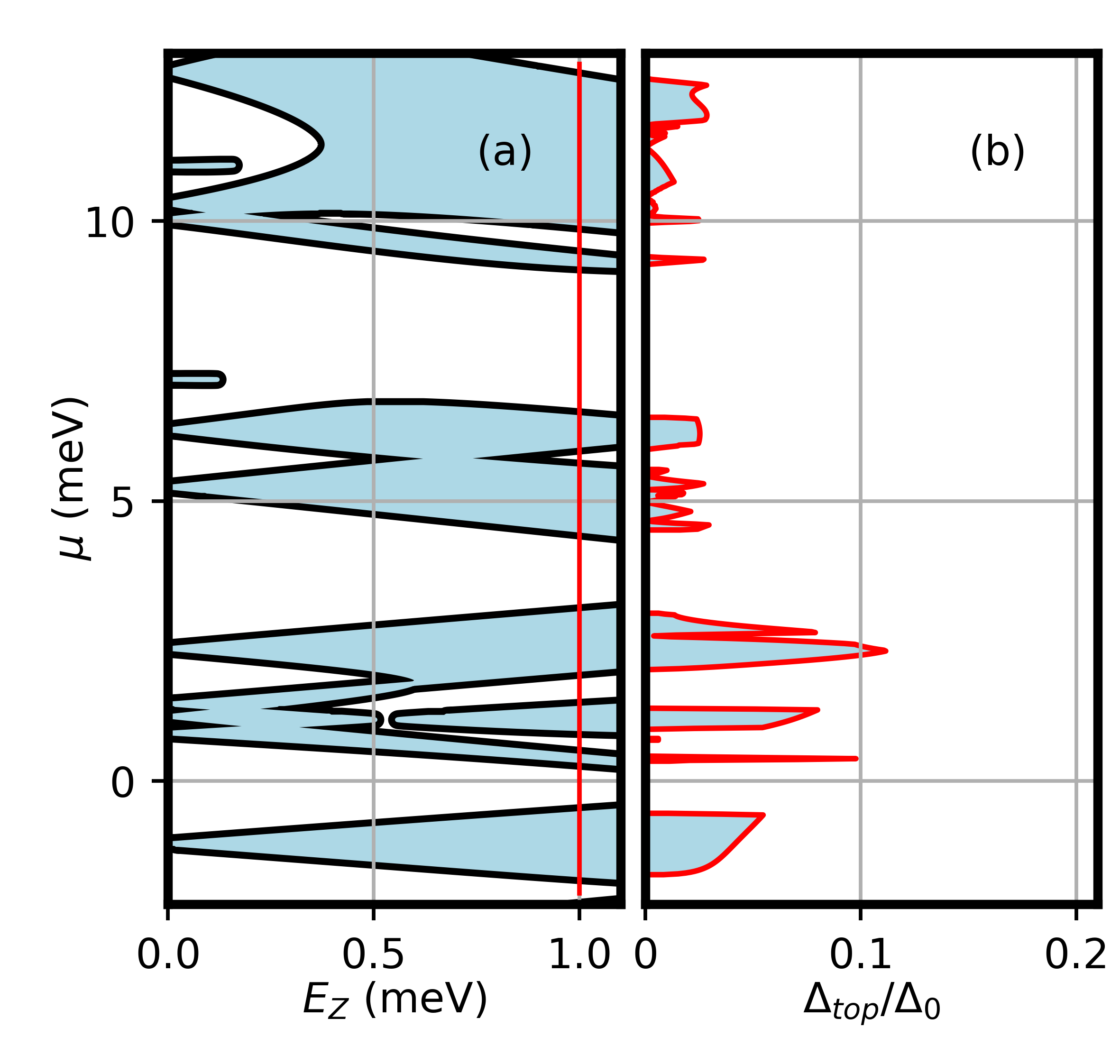}
\end{center}
\vspace{-3mm}
\caption{(a) Low-field phase diagram corresponding to the region $E_Z < 1.1~$meV in Fig. \ref{FIG9}. (b) Topological gap along the cut marked by the red line in (a), which corresponds to a Zeeman field $E_Z=1~$meV. The trivial quasiparticle gap is not shown. Note that the topological gap is characterized by a huge enhancement as compared to the gap of a uniform system in the quantum well regime [$V_J = -40~$meV; see Fig. \ref{FIG3} (c)]. However, the gap values are strongly suppressed in comparison to those of a the modulated system with no phase difference [see Fig. \ref{FIG7} (b)].}
\label{FIG10}
\vspace{-3mm}
\end{figure}

\begin{figure}[t]
\begin{center}
\includegraphics[width=0.48\textwidth]{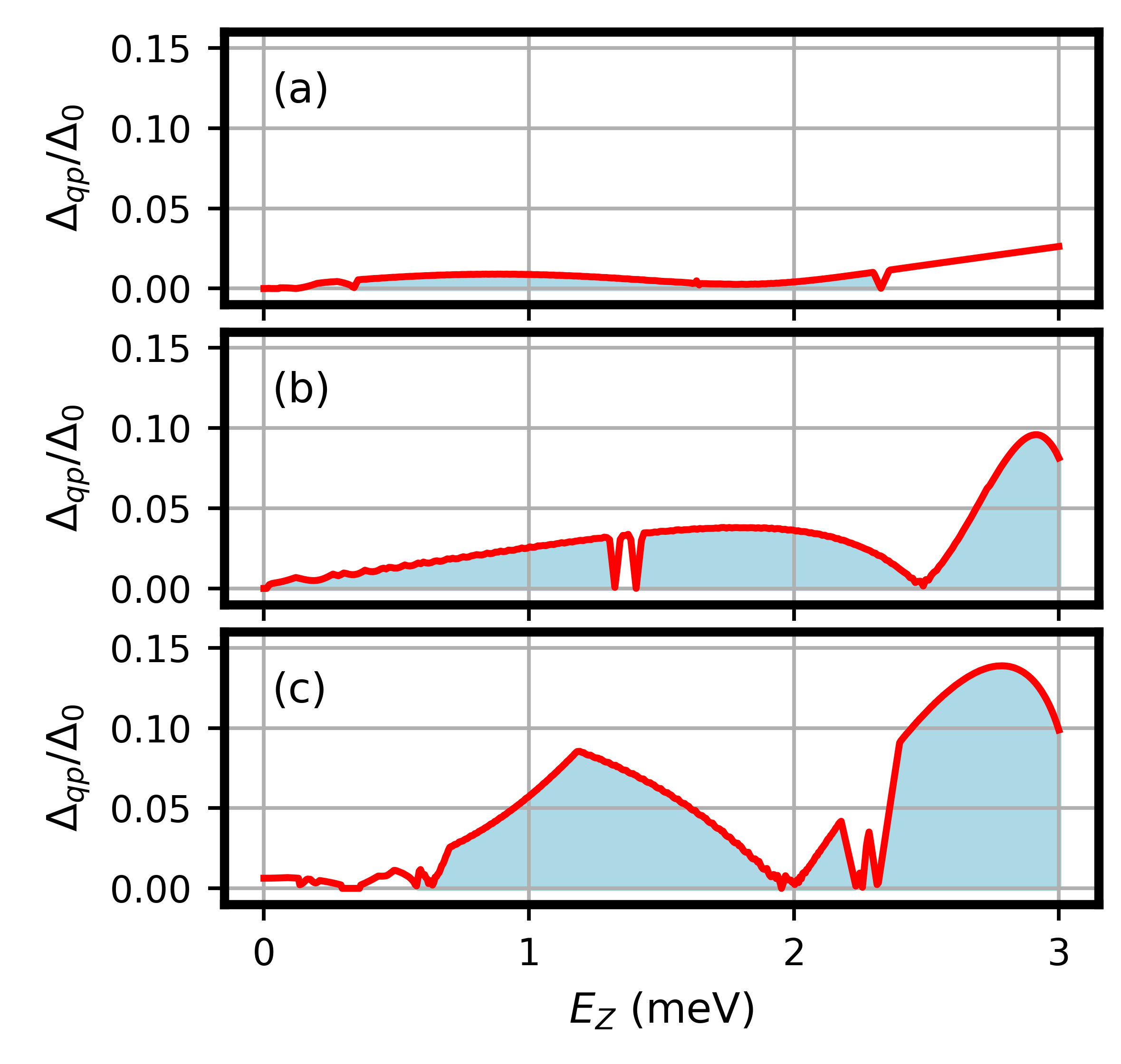}
\end{center}
\vspace{-3mm}
\caption{Quasi-particle gap as a function of $E_Z$ along the horizontal cuts marked by red lines in in Fig. \ref{FIG9}. The chemical potential values are: (a) $\mu = 11.0~\text{meV}$, (b) $\mu = 6.3~\text{meV}$,  and (c) $\mu = 2.7~\text{meV}$. Shading indicates the presence of a topological superconducting phase. The topological gap is massively enhanced in comparison to the corresponding gap of a uniform system (in the quantum well regime) [see Fig. \ref{FIG3}(c)], but is significantly smaller than the gap of the modulated system with no phase difference (Fig. \ref{FIG8}). }
\label{FIG11}
\vspace{-3mm}
\end{figure}

Next, we calculate the size of the quasiparticle gap along representative cuts corresponding to the red lines in Fig. \ref{FIG9}.
It is important to emphasize that the system is deep inside the quantum well regime, $V_J=-40~$meV, and that the corresponding uniform system  would be essentially gapless,  as clearly shown in Fig \ref{FIG3} (c). By contrast, the topological gap characterizing the modulated system with $\phi=\pi$  is finite, although significantly smaller that the corresponding gap in the absence of a phase difference [see Figs. \ref{FIG7} (b)  and \ref{FIG8}].
The results corresponding to the vertical cut are shown in Fig. \ref{FIG10}, with panel (a) reproducing the low-field region of the phase diagram and panel (b) showing the dependence of the topological gap on the chemical potential for $E_Z=1~$meV.
Note that the maximum  value of the topological gap is $\Delta_{\text{top}} \approx 0.1\Delta_o$, with typical values of the order  
 $\Delta_{\text{top}} \approx 0.01 - 0.03 \Delta_o$. By contrast with the system with no phase difference [see Figs. \ref{FIG7} (b)  and \ref{FIG8}], the typical gap tends to decrease with the chemical potential.  This  trend is confirmed by the results in Fig. \ref{FIG11}, which shows the dependence of the quasiparticle gap on the Zeeman field along the horizontal cuts marked by red lines in Fig. \ref{FIG9}. 
 
We conclude that  a modulated structure operated in the quantum well regime ($V_J < 0$) in the presence of a phase difference $\phi=\pi$ exhibits a huge enhancement of the topological gap as compared to a uniform structure operated in the same regime. However, the gap is strongly suppressed as compared to the corresponding values in a modulated structure with no phase difference. More generally, applying a phase difference $\phi \neq 0$ to a modulated structure operated in the quantum well regime reduces the topological gap and, consequently, has a detrimental effect on the robustness of the topological phase. In addition, we have explicitly checked that operating the modulated device in the potential barrier regime, $V_J >0$, leads to effectively loosing the qualitative difference between the modulated and uniform systems. In other words, if $V_J >0$, the experimentally accessible maximum values of the topological gap are similar in modulated and uniform systems. 
 Hence, to take full advantage of the topological enhancement enabled by modulated structures, the device should be operated in the quantum well regime ($V_J<0$) in the absence of a phase difference (i.e., $\phi=0$).

\section{Conclusion} \label{Conclusion}

In summary, we have proposed a planar, periodically modulated  Josephson junction design for realizing topological superconductivity and Majorana zero modes in semiconductor-superconductor hybrid structures. The key feature of the proposed device is represented by the periodic modulation of the junction width, which generates a strong periodic potential within the junction region and induces minibands with strongly renormalized effective parameters. By solving numerically a theoretical model describing the structure, we have shown that modulating the junction width generates several key advantages over non-modulated structures, which ultimately results in an enhancement of topological superconductivity in terms of both accessibility and, most importantly, robustness. More specifically, we have shown that in modulated structures a topological phase characterized by a gap representing a significant fraction of the parent SC gap can be accessed within the low Zeeman field regime by simply tuning the potential in the junction region using, e.g., a top gate. This provides an interesting solution to the main challenge facing the Josephson junction design -- the rather low values of the topological gap and the corresponding fragility of the topological phase -- while preserving its main advantages.
 
An important feature of the ``standard'' JJ design is the substantial expansion of the topological phase within the low Zeeman field regime in the presence of a superconducting phase difference $\phi=\pi$, as illustrated in Fig. \ref{FIG2} (b). Note, however, that an extended topological phase does not automatically translate into practical advantages for realizing robust MZMs. In particular, maximizing the topological gap requires not only finite values of the Zeeman field  that increase with decreasing the width of the junction [see Fig \ref{FIG3}(b) and Fig. \ref{FIG5}], but also fine tuning the potential in the junction region [see Fig. \ref{FIG3}(c)]. In addition, controlling the phase difference can be challenging in a multi-qubit structure. By contrast, our proposed modulated structure is characterized by a topological superconducting phase that covers a significant area of the low-field phase diagram [see figs. \ref{FIG6} and \ref{FIG7}(a)] in the absence of a superconducting phase difference ($\phi =0$). More specifically, the modulated structure should be operated in the quantum well regime, with the junction region having a lower potential than the proximitized regions  ($V_J < 0$), so that bound states can form within the junction. This  can be easily realized using a top gate, since the junction region of the 2DEG is not electrostatically screened by the superconductor. We emphasize that although the low-field topological ``lobes'' in Fig. \ref{FIG6} are similar to features characterizing the phase diagram of ``standard'' multi-band Majorana wires, tuning the chemical potential (or rather the junction potential $V_J$; see Fig. \ref{FIG18}) to access several of these lobes is significantly easier in a planar JJ device as compared to a Majorana wire (or a planar structure with a strip design) because of lack of screening by the parent superconductor.
 
The key advantage of the proposed modulated structure is a substantial enhancement of the topological gap as compared to the optimal experimentally-accessible gap characterizing the uniform, non-modulated structure. This enhancement is due to a stronger effective spin-orbit coupling associated with the  minibands derived from the bound state subbands  that have folded several times in the presence of the effective periodic potential generated by the spatial modulation of the junction width \cite{Woods2020b}. Note that  maximum gap values are obtained in the absence of a superconducting phase difference  across the junction, $\phi = 0$, in the quantum well regime ($V_J < 0$). Also note that a modulated structure with phase difference $\phi = \pi$ operated in the quantum well regime is characterized by a much smaller topological gap [see Figs. \ref{FIG10} (b) and \ref{FIG11}]. However, this represents a dramatic enhancement with respect to the nearly gapless superconducting state associated with the uniform (non-modulated) structure operated in the quantum well regime [see Fig \ref{FIG3}(c)]. In addition, we note that operating the modulated device in the potential barrier regime ($V_J\geq 0$) generates similar gap values as those characterizing the uniform system, regardless of the phase difference $\phi$. 
We emphasize that the example discussed in this work is intended as a proof-of-concept study of the modulation-induced enhancement of the topological gap. In other words, we have not optimized the parameters of the structure. 
 Most likely  the topological gap can be further enhanced by exploiting  higher-order minibands, which can be  reached by making $V_J$  more negative or by increasing $\mu$, and by optimizing the geometric parameters on the junctions (e.g., $\ell$, $L$, $W_1$, $W_2$, etc.). This optimization should be done in synergy with the experimental and engineering efforts to realizing such structures.  On the other hand, we point out that in non-modulated structures the topological gap probably cannot be enhanced significantly due to the intrinsically large Fermi velocity of InAs at even moderate values of the chemical potentials \cite{Pientka2017}. 
 
In addition to enhancing the robustness of the topological phase, the proposed modulated device shares the advantages of the uniform Josephson junction design. In particular, these structures overcome the issues associated with the strong semiconductor-superconductor coupling regime. This is an important property, as most InAs/Al experimental devices, both nanowires and planar structures, appear to be in this regime, which results in a suppression of the effective g-factor and spin-orbit coupling within the proximitized regions.
 Using the junction design removes the engineering challenges associated with tuning the SM-SC coupling. In particular, it simplifies the growth process, since growers can focus solely on creating a clean SM-SC interface, without having to worry about producing barrier layers to reduce the SM-SC coupling. 
 This also may expand the materials combinations that can be explored for building these structures. For example, Pb was recently grown epitaxially on InAs nanowires \cite{Kanne2020}. Importantly Pb is able to withstand a very large magnetic field without SC being destroyed. In the nanowire experiment, however, the SM-SC coupling is clearly in the strong-coupling regime since the bulk gap of the device never closes, even with fields up to 8 $\text{T}$! The topological phase will likely not be achieved in InAs/Pb nanowires, unless the SM-SC coupling is significantly reduced. Provided Pb can be grown cleanly on InAs, our device design overcomes this strong coupling issue. One then may be able to achieve much larger topological gaps than those provided by the  current InAs/Al devices. This of course could significantly increases the robustness of the topological phase against disorder. Note, however, that SCs with larger gaps do not automatically provide a significant enhancement of the topological gap in non-modulated structures with phase difference $\phi=\pi$. In fact, they place stricter requirements on the junction width and increase the Zeeman energy needed to reach a significant topological gap \cite{Pientka2017}. By contrast, our proposed modulated design does not have these limitations.

The most significant potential issue facing the realization of modulated Josephson junction devices concerns the lithography requirements to etch the modulations of the junction geometry. While this may be a challenging engineering and materials growth problem, we estimate that, in the context of a growing interest for nanotechnologies, it is likely that precision lithography will make significant progress  in the coming years, making modulated structures feasible and more attractive. Future theoretical work should address the problem of optimizing the geometric features of the device within the limitations imposed by lithography and investigate the effect of geometric imprecisions associated with lithography, which may represent a significant source of disorder. 
Finally, we note that a number of previous studies have investigated the effects of various types of periodic alterations of the ``basic'' Majorana structures, for both nanowires \cite{Adagideli2014,Levine2017,Escribano2019}  and planar systems \cite{Woods2020b,Laeven2020,Lesser2021}. While significantly different from our proposal, these works have also found  certain benefits of adding  periodic alterations to the uniform structure, which suggests that this type of design deserves further attention. 


~

Author contributions: P.P., T.C., and B.W. contributed equally to this work. B.W. and T.S. conceived of the project idea. T.C. and P.P. performed the numerical calculations with assistance from B.W.. B.W. and T.S.  wrote the manuscript with input from all authors. All authors analysed and discussed the results.

\begin{acknowledgments}
This  work  is  supported  by  NSF Grant No. 2014156
\end{acknowledgments}

\appendix
\renewcommand{\thefigure}{A\arabic{figure}}

\setcounter{figure}{0}

\section{Qualitative effects of modulating the junction width} \label{QEM}

In this Appendix we briefly discuss the effects of the periodic modulation of the junction width at a qualitative level, to highlight the underlying physics. 
Consider first the case of $V_J = 0$, when there is no potential difference between the junction and the proximitized regions. In such a situation, the only difference  between a modulated and a non-modulated structure (with $W_2=W_1$) arises from the changes in $\Delta(x,y)$ and $\widetilde{E}_z(x,y)$ within the region $|y| < W_1/2$ associated with the modulated structure (see Fig. \ref{FIG1}). The relevant question concerns the existence of geometric parameters (i.e., $\ell$, $L$, and $W_2$) consistent with significant deviations from the uniform system. To answer this question, we consider states with energies below the parent gap, $E < \Delta_o$, which are classically forbidden within the proximitized region and have a decay length (approximately) given by 
\begin{equation}
    \xi = 2\sqrt{\frac{\hbar^2}{2m^*} \frac{\mu}{\Delta_o^2 - E^2}}
    = \frac{2 \mu}{k_F \sqrt{\Delta_o^2 - E^2}}, \label{Xi1}
\end{equation}
for $\mu \gg \Delta_o$.  We expect modulation effects to be important if the weight of the wave function within the ``additional'' proximitized regions with $|y| < W_1/2$ (see Fig. \ref{FIG1}) is comparable to the weight of the wave function within the ``standard'' proximitized regions ($|y| > W_1/2$) associated with the uniform system.
 Considering now, as an example, a system with $\mu = 3~\text{meV}$, $E = 0$, and the InAs/Al model parameters given above, Eq. (\ref{Xi1}) yields $\xi \approx 440~\text{nm}$, which is much larger than the values of the geometric parameters considered in this work. This implies that for the uniform system most of the weight associated with the sub-gap state is already within the proximitized regions and that this weight will not change significantly by adding the constrictions. Therefore, we conclude that in the absence of a junction potential $V_J$ modulating the junction width has rather small effects, a conclusion that is confirmed by the numerical calculations. Similar considerations hold when the system is characterized by a positive junction potential, $V_J >0$. Again, the basic reason is that the relevant wave functions undergo negligible changes in the spatial distribution of their spectral weight upon introducing the ``additional'' proximitized regions, which translates into the emergence of a very weak effective periodic potential. 
 We dub the regime characterized by $V_J \geq 0$ as the ``potential barrier'' regime. Based on the above considerations, we conclude that modulating the junction width has weak effects on the low-energy physics of a hybrid structure operating in the potential barrier'' regime.

\begin{figure}[t]
\begin{center}
\includegraphics[width=0.48\textwidth]{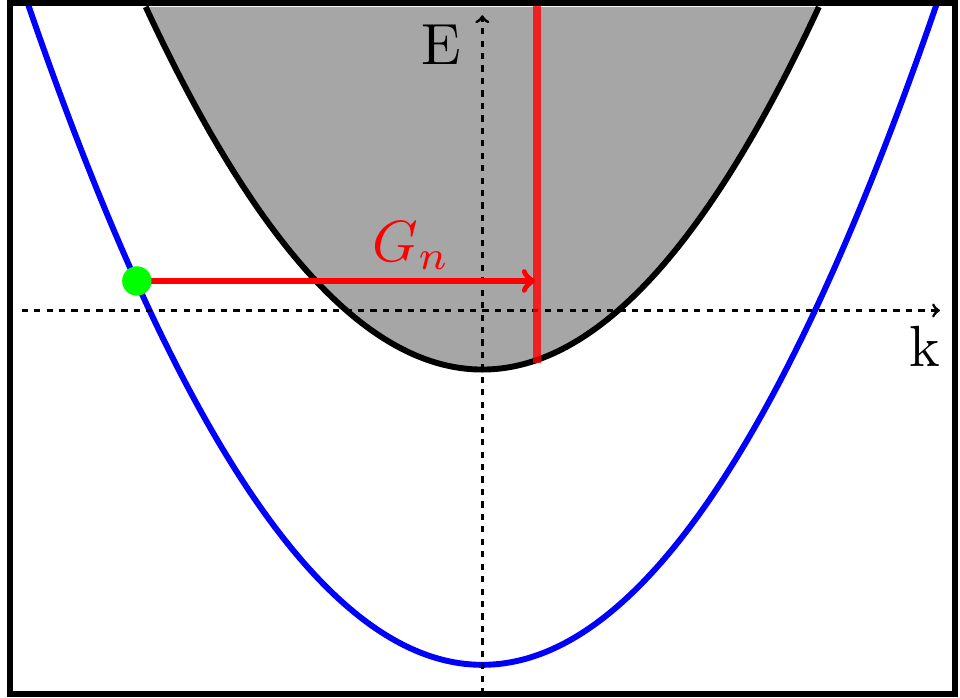}
\end{center}
\vspace{-0.6cm}
\caption{Spectrum of a non-modulated structure without spin-orbit coupling or Zeeman splitting with a sufficiently deep junction potential $V_J<0$ such that a bound state subband (blue line) forms with states heavily localized in the junction region. There also exists a continuum of scatter states (shaded grey region) that extend across the entire device. 
Upon the introduction of modulations, states with $k$'s differing by a reciprocal lattice vector $G_n = \frac{2\pi n}{L}$ couple due to the periodic potential. For example, the bound state indicated by the green dot couples to the continuum of scattering states in the red shaded region. This results in a ``dressed'' bound state with enhanced weight within the proximitized regions.}
\label{FigXX}
\vspace{-1mm}
\end{figure}

Next, we consider the effects of the modulation for a system with $V_J < 0$, when the junction becomes a { quantum well}. The normal spectrum of a non-modulated structure in the absence of spin-orbit coupling and Zeeman splitting is shown in Fig. \ref{FigXX}. 
Assuming that $|V_J|$ is sufficiently large, discrete bound states form within the junction with energies below the continuum of scattering states. Moreover, in the absence of spin-orbit coupling and modulation the $x$ and $y$ components of the problem can be separated. Consequently, the energies of the eigenstates are simply $E_{n,k} = \varepsilon_n + \frac{\hbar^2 k^2}{2 m^*}$, where $k$ is the $x$ component of the momentum and $\varepsilon_n$ is the eigenenergy of the corresponding transverse mode. In addition, the wavefunction of the states take the simple form $\Psi_{n,k}(\mathbf{r}) = \varphi_n(y) \exp(i k x)$, where $\varphi_n$ is the $k$-independent transverse wave function. If we now add superconductivity to the non-modulated system, the bound state bands give rise to an extremely small induced gap, since they are heavily localized within the junction region where $\Delta = 0$. Any topological phase would be extremely fragile in this regime (characterized by $V_J <0$), which we dub the ``quantum well'' regime. Based on these considerations, we conclude that the quantum well regime is basically useless for practical applications involving uniform structures. 

However, introducing periodic modulations of the junction width breaks momentum conservation along the $x$-direction, i.e. $k$ is no longer a good quantum number, but, instead, we have a conserved crystal momentum $q$. This allows states of momentum $k_1$ and $k_2$ to couple through the periodic potential provided $k_2 = k_1 + G_n$, where $G_n$ is a reciprocal lattice vector defined by $G_n = \frac{2 \pi n}{L}$ with $n \in \mathbb{Z}$. As illustrated in Fig. \ref{FigXX}, bound states with energies near the Fermi level can now couple to scattering states with much lower 
momentum generating ``dressed'' bound states that have significantly enhanced weight within the proximitized regions. This strongly alters the previously gapless bound states in the vicinity of the Fermi level, which now acquire a significant induced gap from mixing with (gaped) scattering states. Moreover, the spectrum folds into the first Brillouin zone ($|q| \leq \pi/L$) and the folded subbands associated with ``dressed'' bound states become viable for supporting a topological phase for sufficiently large $E_Z$. In addition, these folded subbands are characterized by renormalized effective parameters (e.g., effective mass and spin-orbit coupling strength), which can result in a significant increase in the (effective) spin-orbit coupling and, consequently, an increase of the topological gap \cite{Woods2020b}. This qualitative picture, which is confirmed by the numerical calculation presented in the Sec. \ref{Results} , captures the key mechanism responsible for the ``topological enhancement'' that characterized our proposed modulated devices.

\begin{figure}[t]
\begin{center}
\includegraphics[width=0.48\textwidth]{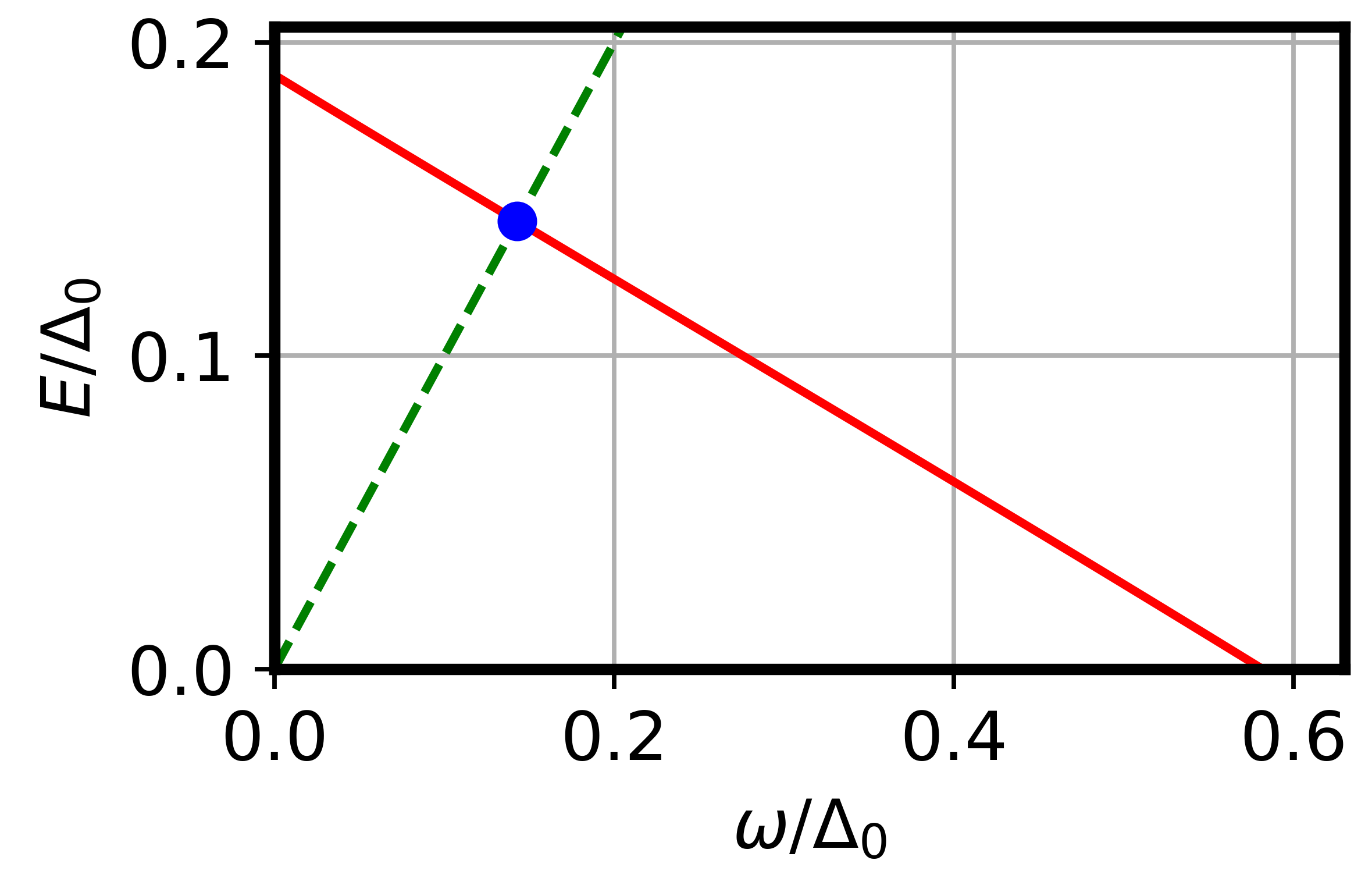}
\end{center}
\vspace{-0.5cm}
\caption{The lowest positive eigenvalue (solid red line) of $\widetilde{\mathcal{H}}(\omega,q)$ monotonically decreases with the energy argument $\omega$ until it passes through zero. The energy of a subgap state is found where the eigenvalue and $E = \omega$ (dashed green line) curves intersect (blue dot). System parameters are given by $\mu = 1$ meV, $E_Z = 1$ meV, $\phi = \pi$, with the same geometric parameters as Fig. \ref{FIG2}.}
\label{FIG17A}
\vspace{-1mm}
\end{figure}

\section{Calculation of the topological gap} \label{Iter}

In this Appendix, we explain how we use a self-consistent numerical procedure to calculate the energy of eigenstates below the bulk superconducting gap $\Delta_o$ and find the topological gap when the system is in the topological phase. 

The energy E of a subgap state with crystal momentum $q$ corresponds to a pole of the reduced Green's function $G_J(\omega,q)$ in Eq. (\ref{GreensJunc}) such that $|\omega| < \Delta_o$. Finding these poles is equivalent to solving the eigenvalue equation,
\begin{equation}
\widetilde{\mathcal{H}}_{J}^{\prime}( E,q)\psi = E\psi, \label{SchSC}
\end{equation}
where
\begin{equation}
    \widetilde{\mathcal{H}}_{J}^{\prime}( \omega,q) = 
    \widetilde{\mathcal{H}}_J(q) 
    + \Sigma_{SC}(\omega,q).
\end{equation}
Note that the energy dependence of $\widetilde{\mathcal{H}}_{J}^{\prime}$ requires that Eq. (\ref{SchSC}) be solved self-consistently, namely the energy argument of $\widetilde{\mathcal{H}}_{J}^{\prime}$ must equal one of the eigenvalues of $\widetilde{\mathcal{H}}_{J}^{\prime}$. The lowest positive eigenvalue of $\widetilde{\mathcal{H}}_{J}^{\prime}$ is shown as the red line in Fig. \ref{FIG17A} for arbitrary system parameters of a uniform junction in the topological phase.
The energy of a subgap state satisfying Eq. (\ref{SchSC}) is given by the intersection of the eigenvalue curve with the curve $E = \omega$ (green dashed line) in Fig. \ref{FIG17A}. Note that the eigenvalue monotonically decreasing with an increasing value of the energy argument until the eigenvalue passes through zero near $\omega \approx 0.6 \Delta_o$. Additionally note that this is the generic behavior we observe for all system parameters. This enables us to use an efficient root-finding algorithm that iteratively reduces the energy interval in which we can find the intersection corresponding to the energy of a subgap state. We continue the iteration process until the length of the energy interval is below a small tolerance.

\begin{figure}[t]
\begin{center}
\includegraphics[width=0.48\textwidth]{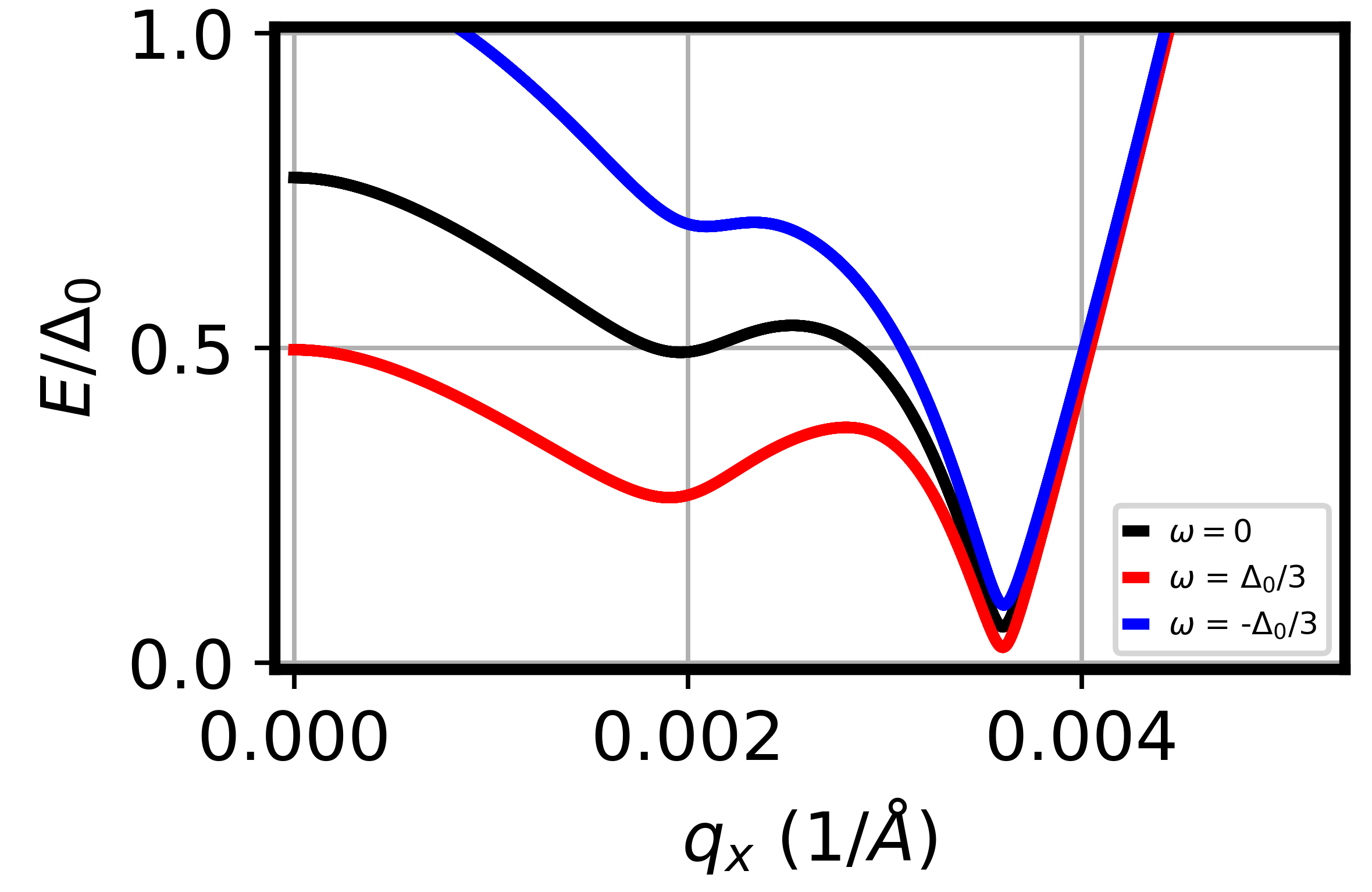}
\end{center}
\vspace{-0.5cm}
\caption{Lowest positive eigenvalue of $\widetilde{\mathcal{H}}_{J}^{\prime}(\omega,q)$ as a function of $q_x = q/L$ for three different values of the energy argument $\omega$. While the energy argument $\omega$ significantly affects the eigenvalue curve, we find that the minimum eigenvalue always occurs at the same $q$ value independent of the energy argument $\omega$.}
\label{FIG18A}
\vspace{-1mm}
\end{figure}

To then calculate the value of topological gap, we must find the minimum subgap energy which can occur at any value of crystal momentum $q$. A brute force approach would require scanning $q$ and performing the self-consistency algorithm described above for each value. Fortunately, we generically find the minimum  eigenvalue of $\widetilde{\mathcal{H}}_{J}^{\prime}(\omega,q)$ occurs at the same $q$ independent of the energy argument $\omega$. This is illustrated in Fig. \ref{FIG18A}, where the lowest eigenvalue of $\widetilde{\mathcal{H}}_{J}^{\prime}$ is plotted for three different energy argument values. We therefore first find the $q = q_{min}$ yielding the minimum eigenvalue of $\widetilde{\mathcal{H}}_{J}^{\prime}(\omega = 0,q)$. Next we perform the self-consistent algorithm described above for $\widetilde{\mathcal{H}}_{J}^{\prime}(\omega,q = q_{min})$, which finally then yields the topological gap.


\maketitle
\section{Nearly gapless topological phase at $\mathbf{q\neq 0,\pi}$} 

\begin{figure}[t]
\begin{center}
\includegraphics[width=0.48\textwidth]{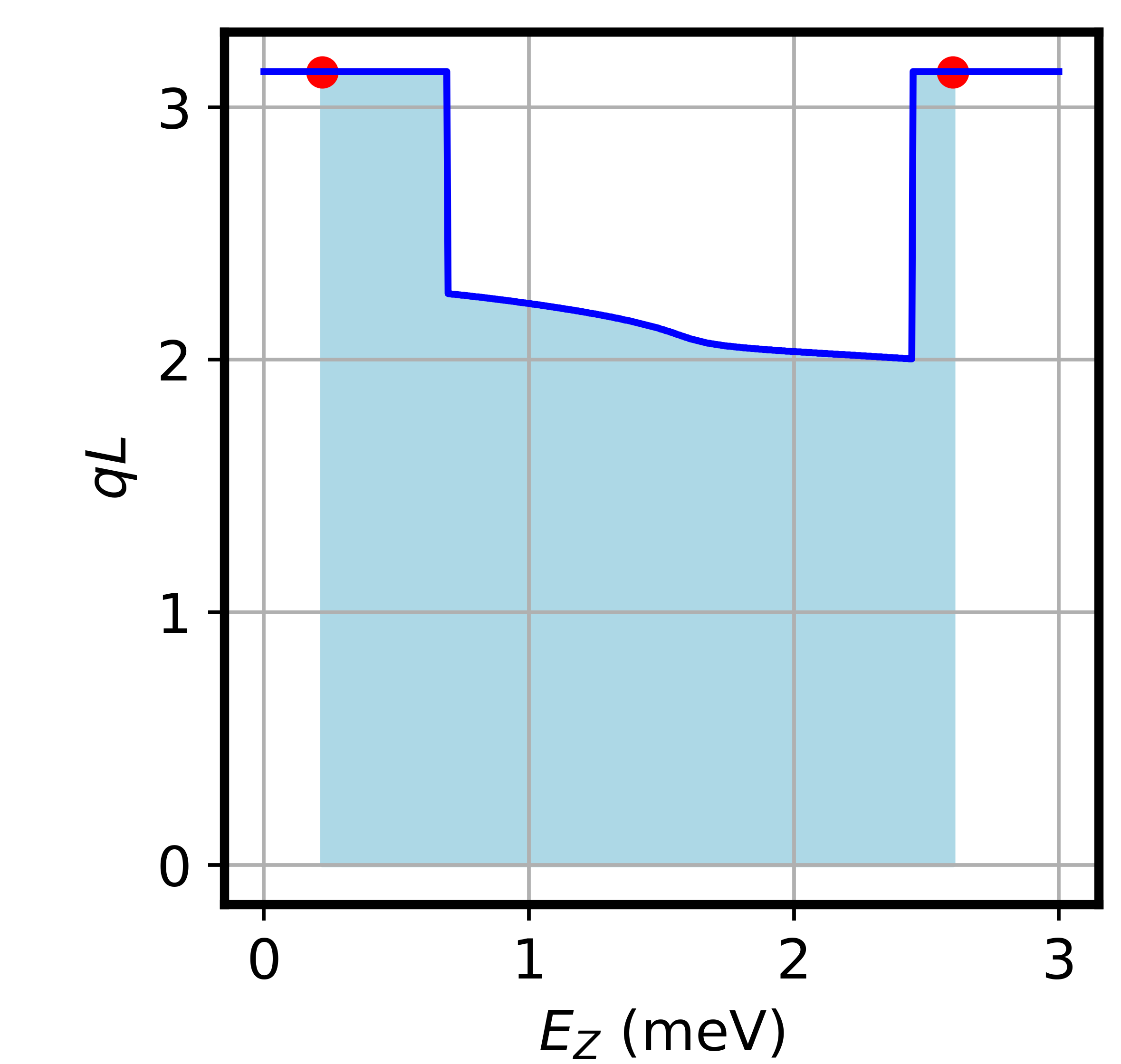}
\end{center}
\vspace{-0.5cm}
\caption{Crystal momentum with minimum quasi-particle gap $q_{min}$ as a function of Zeeman energy $E_z$ for the same system as Fig. \ref{FIG8} (b) of the main text. Shaded blue indicates that the system is in the topological phase. Topological phase transitions correspond to bulk gap closures at $q_{min} = \pi$ and are indicated by red dots. The small topological gap region near $E_z = 2.2~\text{meV}$ (see Fig. \ref{FIG8}) is characterized by $q_{min} \neq 0, \pi$, indicating that we do not have a topological phase transition near this $E_z$ value.}
\label{FIG15A}
\vspace{-1mm}
\end{figure}

We find that our proposed modulated JJ system is sometimes characterized by a topological phase with a nearly vanishing topological gap. For example, in Fig. \ref{FIG8} (b) of the main text the topological gap appears to vanish near $E_z = 2.2~\text{meV}$. Naively, this may initially be attributed to a topological phase transition. This is not the case, however. If we were to zoom in Fig. \ref{FIG8} (b) near $E_z = 2.2~\text{meV}$, we would find a small, but non-zero, topological gap. Additionally, recall that the topological phase transitions in 1D class D systems occur due to bulk gap closures at $q = 0,\pi$ \cite{Chiu2016}. We find, however, that the crystal momentum $q_{min}$ responsible for the topological gap near this value of $E_z$ is not at the zone boundary, i.e. $q_{min} \neq 0,\pi$. This is illustrated in Fig. \ref{FIG15A}, which shows $q_{min}$ as a function of Zeeman energy for the same system as Fig. \ref{FIG8} (b). Therefore, even if the topological gap were to actually vanish near $E_z = 2.2~\text{meV}$ instead of being a small value, this would not be an indication of a topological phase transition. Rather, our modulated JJ system would then be a gapless topological superconductor. 

\section{Investigating the role of junction asymmetry} 

\begin{figure}[t]
\begin{center}
\includegraphics[width=0.48\textwidth]{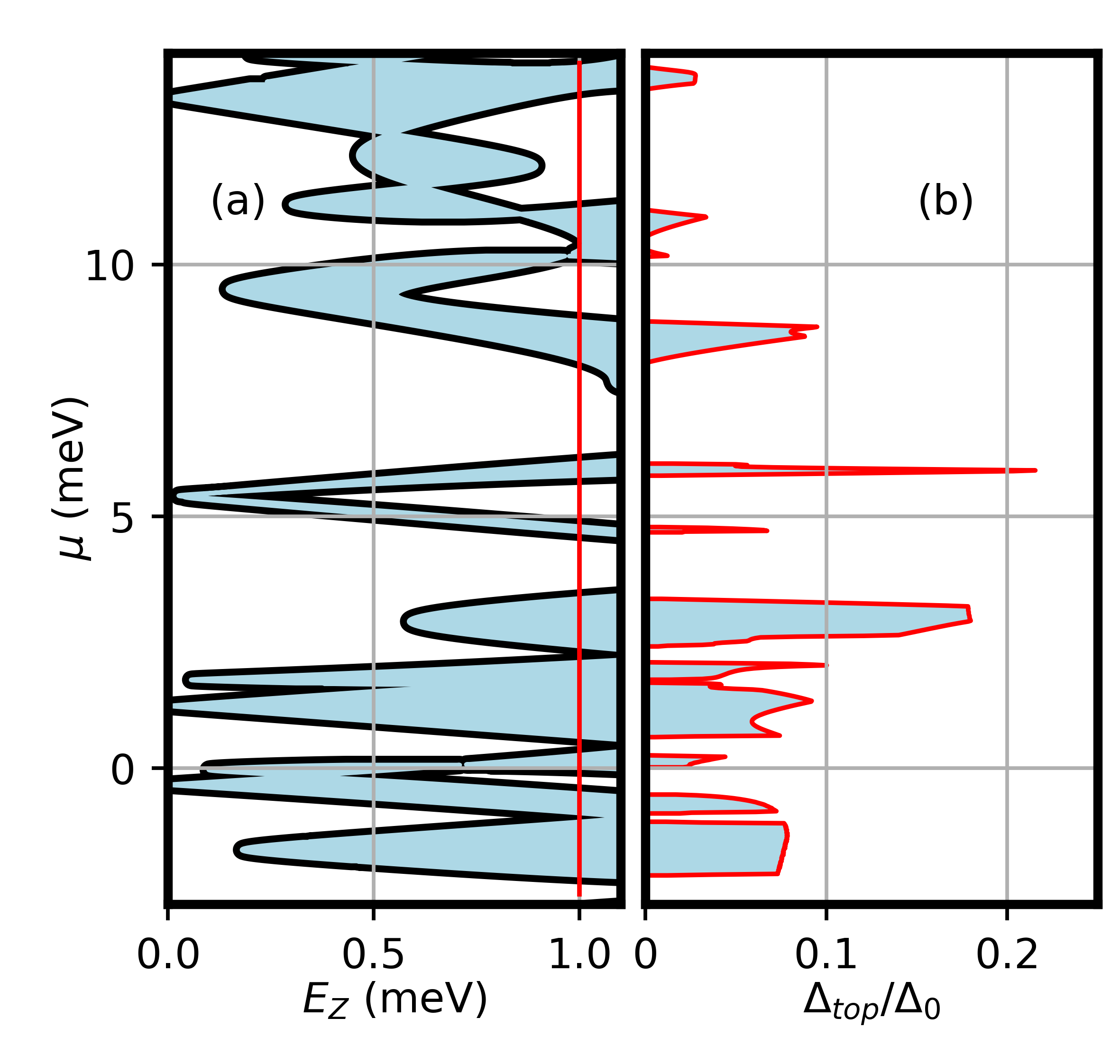}
\end{center}
\vspace{-0.5cm}
\caption{Phase diagram and topological gap plot for the asymmetric junction with SC phase difference $\phi$ =  $\pi$. (a) Phase diagram at low Zeeman energy regime. (b) Topological gap calculated along the vertical line cut taken at $E_Z = 1 $ meV as shown in (a).}
\label{FIG15}
\vspace{-1mm}
\end{figure}

The geometric parameters of all of the modulated JJs that we studied in the main text were chosen such that the system had reflection symmetry about the $x$-axis. A natural question is what effect does breaking the reflection symmetry have on the topological phase diagram and the topological gap? We investigate this question by studying one asymmetric modulated JJ with geometric parameters $W_1 = 100~\text{nm}$, $W_2 = 20~\text{nm}$, $L = 60~\text{nm}$, and $\ell = 20~\text{nm}$, and $w = 0$. Note that modulated structure corresponding to Figs. \ref{FIG6}-\ref{FIG11} in the main text have the same geometric parameters except $w \neq 0$. The topological phase diagram with $\phi = \pi$ is shown in Fig. \ref{FIG15} (a), and the topological gap for fixed $E_z = 1~\text{meV}$ is shown in \ref{FIG15} (b). Compared to the corresponding phase diagram in Fig. \ref{FIG10} of the symmetric system, the results are qualitatively similar. It does appear that the topological gap may be slightly enhanced in the asymmetric system compared to the symmetric structure, although the topological gap still remains below $0.2\Delta_o$ except for the narrow region near $\mu = 6~\text{meV}$. At the same time, however, the topological regions of the asymmetric structure in Fig. \ref{FIG15} (a) emerge at slightly larger $E_z$ on average when compared to the symmetric results in Fig. \ref{FIG10} (a). Considering the fact that the effects of introducing asymmetry in this structure are small, we conclude that the presence or absence of reflection symmetry does not likely play a crucial role.

\vspace{9cm} 

\end{document}